\documentclass[a4paper,11pt]{article}
\usepackage{jheppub} 
\usepackage{dcolumn}
\usepackage{bm}
\usepackage{amsmath}
\usepackage{amsfonts}
\usepackage{amssymb}
\usepackage{graphicx,subfigure}
\usepackage{color}
\usepackage{latexsym}
\usepackage{slashed} 
\usepackage{mathrsfs}
\usepackage{multirow}
\usepackage{epsfig,psfrag}
\usepackage[scientific-notation=true]{siunitx} 
\usepackage[utf8]{inputenc} 

\graphicspath{{fig/}} 

\makeatletter
\newcommand\figcaption{\def\@captype{figure}\caption} \newcommand\tabcaption{\def\@captype{table}\caption}
\makeatother

\allowdisplaybreaks 

\begin{document}
\title{CEPC Precision of Electroweak Oblique Parameters and Weakly Interacting Dark Matter: the Fermionic Case}
\author[a]{Chengfeng Cai,}
\author[b]{Zhao-Huan Yu,}
\author[a]{and Hong-Hao Zhang\footnote{Corresponding author}}

\affiliation[a]{School of Physics, Sun Yat-Sen University, Guangzhou 510275, China}
\affiliation[b]{ARC Centre of Excellence for Particle Physics at the Terascale,
School of Physics, The University of Melbourne, Victoria 3010, Australia}

\emailAdd{caichf@mail2.sysu.edu.cn}
\emailAdd{zhao-huan.yu@unimelb.edu.au}
\emailAdd{zhh98@mail.sysu.edu.cn}

\abstract{Future electroweak precision measurements in the Circular Electron Positron Collider (CEPC) project would significantly improve the precision of electroweak oblique parameters. We evaluate the expected precision through global fits, and study the corresponding sensitivity to weakly interacting fermionic dark matter. Three models with electroweak multiplets in the dark sector are investigated as illuminating examples. We find that the CEPC data can probe up to TeV scales and explore some regions where direct detection cannot reach, especially when the models respect the custodial symmetry.}

\maketitle

\section{Introduction}

The standard model (SM) has achieved a great success in explaining how the physics happens at subatomic scales.
However, there are several puzzles that have not yet been solved.
One of the most famous puzzles is dark matter (DM), which makes up most of the matter component in the Universe according to cosmological and astrophysical observations (See, for example,~\cite{Jungman:1995df,Bertone:2004pz,Feng:2010gw} for reviews).
It is strongly suggested that DM might be some kind of weakly interacting massive particles (WIMPs), as they can give a desired relic abundance through thermal production in the early Universe.

WIMPs are typically introduced in the SM extensions for solving the gauge hierarchy problem, such as supersymmetric~\cite{Goldberg:1983nd,Ellis:1983ew} and extra dimensional~\cite{Servant:2002aq,Cheng:2002ej} models.
The common ingredient in these models for explaining dark matter is that WIMPs appear as colorless electroweak (EW) multiplets whose electrically neutral components serve as DM candidates.
Therefore, it would be more general to consider WIMP models with a dark sector consisting of EW multiplets.
The simplest choice is to introduce a multiplet in a nontrivial $\mathrm{SU}(2)_\mathrm{L}$ representation and hence the model is called minimal dark matter~\cite{Cirelli:2005uq,Cirelli:2007xd,Cirelli:2009uv,Hambye:2009pw,Cai:2012kt,Ostdiek:2015aga,Cai:2015kpa}.
The next-to-minimal construction is to make use of two $\mathrm{SU}(2)_\mathrm{L}$ representations, leading to a wider theoretical landscape and a richer phenomenology~\cite{Mahbubani:2005pt,D'Eramo:2007ga,Enberg:2007rp,Cohen:2011ec,Fischer:2013hwa,Cheung:2013dua,Dedes:2014hga,Calibbi:2015nha,Freitas:2015hsa,Fedderke:2015txa,Yaguna:2015mva,Tait:2016qbg,Horiuchi:2016tqw,Banerjee:2016hsk,Kakizaki:2016dza}.

The introduction of extra EW multiplets will affect EW precision observables via loop corrections. As a result, accurate measurements of these observables may give some hints to this kind of new physics. Currently, the most precise results come from the measurements in LEP, SLC, Tevatron, and LHC experiments. In particular, the discovery of the Higgs boson~\cite{Aad:2012tfa,Chatrchyan:2012xdj} fixes its mass, the final free parameter in the SM, and hence greatly improves the global electroweak fit~\cite{Agashe:2014kda,Ciuchini:2013pca,Baak:2014ora,deBlas:2016ojx}.
The recently proposed circular electron-positron collider (CEPC) in China~\cite{CEPC-SPPCStudyGroup:2015csa} would mainly serves as a Higgs factory, aiming at precision measurements of the Higgs physics. Besides, it could also carry out new scans around the $Z$ pole and the $W^+W^-$ threshold with a high luminosity, leading to essential improvements for the measurements of most EW precision observables. This would provide an excellent opportunity to indirectly probe WIMP models.
Recent works on the CEPC sensitivity to new physics through EW precision measurements include studies
on the anomalous $hhh$ and $htt$ couplings through the $e^+e^-\to Zh$ measurement~\cite{McCullough:2013rea,Shen:2015pha,Huang:2015izx,Kobakhidze:2016mfx},
natural supersymmetry~\cite{Fan:2014axa},
the anomalous $hZ\gamma$ and $h\gamma\gamma$ couplings through the $e^+e^-\to h\gamma$ measurement~\cite{Cao:2015iua,Hu:2014eia},
the anomalous $Zbb$ coupling~\cite{Gori:2015nqa},
WIMP models through the $e^+e^-\to f\bar{f}$ measurements~\cite{Harigaya:2015yaa,Cao:2016qgc},
effective operators~\cite{Fedderke:2015txa,Ge:2016zro},
and so on.

In global fits of EW precision observables, the oblique parameters $S$, $T$, and $U$~\cite{Peskin:1990zt,Peskin:1991sw} are often introduced to characterize the general effects of new EW particles that do not directly couple to SM fermions. Since a $Z_2$ symmetry is typically imposed to forbid DM couplings to SM fermions in order to stabilize DM candidates, these parameters are extremely suitable for exploring WIMP models. The CEPC precision of the oblique parameters has been estimated in Ref.~\cite{Fan:2014vta} and been incorporated into the preliminary conceptual design report of CEPC~\cite{CEPC-SPPCStudyGroup:2015csa}. However, the results are only obtained under the assumptions of $U=0$, which is appropriate because $U$ is much smaller than $T$ for a wide class of new physics models. In general, this is not true. For instance, $U$ can be large if there are anomalous triple gauge couplings~\cite{Altarelli:1990zd,Altarelli:1991fk}. For this reason, a global fit with free $U$ could be also useful.

In this work, we will at first perform a global fit to derive the CEPC precision of EW oblique parameters, based on the expected uncertainties of EW precision observables in CEPC measurements with the latest updated results. In order to fulfill different needs, we will study the case where $S$, $T$, and $U$ are all free parameters, as well as the cases where some of them are fixed to zero. We will then use the fit results to investigate the CEPC capability for testing WIMP models.
Only the fermionic DM case will be discussed hereafter. We leave the discussion of the scalar DM case to a future paper.

Renormalizable couplings between fermions and the Higgs must be Yukawa couplings, which require that there are two fermionic multiplets belonging to two $\mathrm{SU}(2)_\mathrm{L}$ representations whose dimensions differ by one, because the SM Higgs field is a doublet.
We introduce a $Z_2$ symmetry to protect DM from decaying.
If there is just one vector-like fermionic multiplet in the dark sector, it cannot couple to the Higgs, because it cannot have Yukawa couplings with SM fermions due to the $Z_2$ symmetry.\footnote{Note that in minimal dark matter models, the $Z_2$ symmetry is not necessary and the DM is accidentally stable, as long as the multiplet lives in a representation with a sufficient high dimension~\cite{Cirelli:2005uq}. But we do not consider such high dimensional representations in this paper.}
In this case, the components of the multiplet are degenerate in mass at tree level, because there is no mass contribution from EW symmetry breaking.
However, a vector-like fermionic or scalar $\mathrm{SU}(2)_\mathrm{L}$ multiplet cannot contribute to $S$, $T$, or $U$ if its components have exactly degenerate masses~\cite{Zhang:2006de,Zhang:2006vt}. Consequently, that kind of fermionic minimal WIMP models (with only one dark sector multiplet) predicts vanishing EW oblique parameters at leading order. Of course, it is \textit{not} the case we plan to discuss in this paper.

The simplest way to avoid it is to consider the kind of fermionic WIMP models with two types of $\mathrm{SU}(2)_\mathrm{L}$ multiplets whose dimensions differ by one. The two types of fermion multiplets couple to each other by Yukawa interactions with the SM Higgs, and the fermion components will be splitted when the Higgs field develops a vacuum expectation value (VEV).
This kind of models is sensitive to electroweak precision measurements and thus will be detectable at the future high precision electron-positron colliders such as the CEPC.
Based on this observation, below we will study three WIMP models with fermionic $\mathrm{SU}(2)_\mathrm{L}$ multiplets in vector-like representations as illuminating examples:
\begin{itemize}
\item Singlet-Doublet Fermionic Dark Matter (SDFDM): one singlet Weyl spinor and two doublet Weyl spinors~\cite{Mahbubani:2005pt,D'Eramo:2007ga,Enberg:2007rp,Cohen:2011ec,Calibbi:2015nha};
\item Doublet-Triplet Fermionic Dark Matter (DTFDM):  two doublet Weyl spinors and one triplet Weyl spinor~\cite{Dedes:2014hga};
\item Triplet-Quadruplet Fermionic Dark Matter (TQFDM): one triplet Weyl spinor and two quadruplet Weyl spinors~\cite{Tait:2016qbg}.
\end{itemize}

In each model, a $Z_2$ symmetry is imposed and the dark sector fermionic fields are odd under this symmetry for stabilizing the DM candidate.
In order to cancel gauge anomalies, the spinor in the odd-dimensional representation has no hypercharge, while the two spinor in the  even-dimensional representation have opposite hypercharges, whose values are assigned to be $\pm 1/2$ for allowing Yukawa couplings with the SM Higgs field.
Once the Higgs obtains a non-zero VEV, the neutral components of the multiplets will be mixed up and the lightest mass eigenstate, which is a Majorana fermion, would be a DM candidate.

These models are quite predictive, as there are just four new parameters, two mass parameters and two Yukawa couplings.
In addition, they may be treated as subsets of more sophisticated models, sharing some of the related phenomenology. For instance, the SDFDM model may correspond to the bino-Higgsino sector in the MSSM or the singlino-Higgsino sector in the NMSSM, while the DTFDM model may correspond to the Higgsino-wino sector in the MSSM.

The paper is organized as follows.
In Section~\ref{sec:STU}, we carry out a global fit to obtain the expected CEPC precision of EW oblique parameters $S$, $T$, and $U$.
In Sections~\ref{sec:SDFDM}, \ref{sec:DTFDM}, and \ref{sec:TQFDM}, we  study the SDFDM, DTFDM, and TQFDM models in details, respectively. We estimate the expected constraints on these models from the oblique parameters with the CEPC precision, as well as current constraints from DM direct detection experiments for comparison.
Section~\ref{sec:concl} gives our conclusions and additional discussions. Appendix \ref{app:DM_DD} supplements formulas for computing the spin-independent and the spin-dependent DM-nuclei scattering cross sections.

\section{CEPC Precision of Electroweak Oblique Parameters}
\label{sec:STU}

In this section, we perform a global fit to estimate the CEPC precision of EW oblique parameters $S$, $T$, and $U$, based on the expected improvements in future measurements.

\subsection{Electroweak Oblique Parameters}

EW radiative corrections can be categorized into two classes, ``direct'' corrections (vertex, box, and bremsstrahlung corrections) and ``oblique'' corrections (gauge boson propagator corrections). While the former is process-specific, the latter is not: oblique corrections are universal, as they appear in any process mediated by EW gauge bosons.
Oblique corrections can be well organized in the Kennedy-Lynn formalism~\cite{Kennedy:1988sn}, which uses an effective lagrangian to incorporate gauge boson vacuum polarization diagrams into a few running couplings. Following this model-independent formalism, Peskin and Takeuchi introduced EW oblique parameters $S$, $T$, and $U$ to describe new physics contributions through EW oblique corrections~\cite{Peskin:1990zt,Peskin:1991sw}.

The definitions of these parameters are
\begin{eqnarray}
S&=&16\pi[\Pi'_{33}(0)-\Pi'_{3Q}(0)],\\
T&=&\frac{4\pi}{s_\mathrm{W}^2c_\mathrm{W}^2m_Z^2}[\Pi_{11}(0)-\Pi_{33}(0)],\label{eq:def:T}\\
U&=&16\pi[\Pi'_{11}(0)-\Pi'_{33}(0)].\label{eq:def:U}
\end{eqnarray}
where $\Pi'_{IJ}(0)\equiv\partial\Pi_{IJ}(p^2)/\partial p^2|_{p^2=0}$. $\Pi_{11}$, $\Pi_{33}$, and $\Pi_{3Q}$ are related to the $g_{\mu\nu}$ coefficients of the vacuum polarization amplitudes of EW gauge bosons contributed by new physics:
\begin{eqnarray}
\Pi_{WW}(p^2)&=&\frac{e^2}{s_\mathrm{W}^2}\Pi_{11}(p^2),\\
\Pi_{ZZ}(p^2)&=&\frac{e^2}{s_\mathrm{W}^2c_\mathrm{W}^2}[\Pi_{33}(p^2)-2s_\mathrm{W}^2\Pi_{3Q}(p^2)+s_\mathrm{W}^4\Pi_{QQ}(p^2)],\\
\Pi_{ZA}(p^2)&=&\frac{e^2}{s_\mathrm{W}c_\mathrm{W}}[\Pi_{3Q}(p^2)-s_\mathrm{W}^2\Pi_{QQ}(p^2)],\\
\Pi_{AA}(p^2)&=&e^2\Pi_{QQ}(p^2).
\end{eqnarray}
Here $s_\mathrm{W}\equiv\sin\theta_\mathrm{W}$ and $c_\mathrm{W}\equiv\cos\theta_\mathrm{W}$ with $\theta_\mathrm{W}$ denoting the Weinberg angle.
$S$, $T$, and $U$ are dimensionless by definition.
Since only new physics contributions to these parameters are taken into account, SM corresponds to $S=T=U=0$.

From definitions~\eqref{eq:def:T} and \eqref{eq:def:U} one can easily see that compared with $T$, $U$ is typically suppressed by a factor of $m_Z^2/m_\mathrm{new}^2$, where $m_\mathrm{new}$ represents the mass scale of a new EW sector~\cite{Peskin:1990zt}.
This point can also be understood in the context of effective field theory as follows. $S$ and $T$ correspond to dimension-6 operators ${H^\dag }W_{\mu \nu }^a{\sigma ^a}H{B^{\mu \nu }}$ and ${H^\dag }({D_\mu }H)({D^\mu }H )^\dag H$, respectively, while the operator contributing to $U$ in the lowest order is a dimension-8 operator ${H^\dag }W_{\mu \nu }^a{\sigma ^a}H{H^\dag }{W^{b\mu \nu }}{\sigma ^b}H$ (See, for instance, Ref.~\cite{Han:2008es} for a review).
Thus, many typical new physics models predict $U\ll T$.
This is the reason why the assumption $U=0$ is often adopted in global EW fits.
The operator ${H^\dag }W_{\mu \nu }^a{\sigma ^a}H{B^{\mu \nu }}$ means that $S$ is related to the $\mathrm{U}(1)_\mathrm{Y}$ gauge field. Actually, the contribution to $S$ from a vector-like fermionic multiplet or a scalar multiplet is proportional to its hypercharge~\cite{Zhang:2006de,Zhang:2006vt}.

Experimental results show that the observable $\rho=m_W^2/(m_Z^2 c_\mathrm{W}^2)$ is extremely close to one~\cite{Agashe:2014kda}. In the SM, $\rho=1$ is an exact relation at tree level. It has been argued that this relation naturally holds up to EW radiative corrections if the Higgs sector has an $\mathrm{SU}(2)_\mathrm{R}$ global symmetry. Once the Higgs field obtains a nonzero VEV, the $\mathrm{SU}(2)_\mathrm{L}\times \mathrm{SU}(2)_\mathrm{R}$ global symmetry spontaneously breaks down to the unbroken $\mathrm{SU}(2)_\mathrm{L+R}$ symmetry, \textit{i.e.} the so-called custodial symmetry~\cite{Sikivie:1980hm}. Under such a symmetry the $\mathrm{SU}(2)_\mathrm{L}$ gauge bosons $W^a_\mu$ ($a=1,2,3$) transform as a triplet and thus acquire the same mass. This property and the fact that a $\mathrm{U}(1)_\mathrm{em}$ gauge symmetry is unbroken result in $\rho=1$.
In the language of oblique parameters, the deviation of the $\rho$ parameter from one is $\alpha T$~\cite{Peskin:1991sw,Zhang:2009rm}, and the existence of a custodial symmetry will lead to $T=U=0$.

In a global EW fit, each EW precision observable $O$ can be split into two parts: $O=O_\mathrm{SM}+\Delta O(S,T,U)$. The SM contribution $O_\mathrm{SM}$ should be computed as accurately as possible, and the new physics contribution $\Delta O(S,T,U)$ is a functions of the three oblique parameters. In fact, any $\Delta O(S,T,U)$ should be proportional to one of the following functions~\cite{Ciuchini:2013pca}:
\begin{eqnarray}
{F_1}(S,T,U) &=& S - 2c_{\mathrm{W}}^2T - \frac{{c_{\mathrm{W}}^2 - s_{\mathrm{W}}^2}}{{2s_{\mathrm{W}}^2}}U,
\label{eq:F1}\\
{F_2}(S,T) &=& S - 4s_{\mathrm{W}}^2c_{\mathrm{W}}^2T,
\label{eq:F2}\\
{F_3}(S,T) &=&  - 10(3 - 8s_{\mathrm{W}}^2)S + (63 - 126s_{\mathrm{W}}^2 - 40s_{\mathrm{W}}^4)T.
\label{eq:F3}
\end{eqnarray}

\subsection{Electroweak Precision Observables}

In light of the strategy for studying the prospect of EW precision tests  used in Refs.~\cite{Baak:2014ora,Fan:2014vta}, we adopt a simplified set of precision observables:
\begin{enumerate}
\item ${\alpha _{\mathrm{s}}}(m_Z^2)$, the strong coupling constant at the $Z$ pole;
\item $\Delta \alpha _{{\mathrm{had}}}^{(5)}(m_Z^2)$, the quark sector contribution (without the top quark) to the running of the QED coupling $\alpha$ at the $Z$ pole;
\item ${m_Z}$, the $Z$ boson pole mass;
\item ${m_t}$, the top quark pole mass;
\item ${m_h}$, the Higgs boson pole mass;
\item ${m_W}$, the $W$ boson pole mass;
\item ${\sin ^2}\theta _{{\mathrm{eff}}}^\ell $, the effective weak mixing angle for the $Z\ell\ell$ coupling;
\item ${\Gamma _Z}$, the $Z$ boson decay width.
\end{enumerate}
In our global fit, the first five observables, as well as $S$, $T$, and $U$, are treated as free parameters. The SM predictions of the remaining three, $m_W^\mathrm{SM}$, $({\sin ^2}\theta _{{\mathrm{eff}}}^\ell )^\mathrm{SM}$, and $\Gamma _Z^\mathrm{SM}$, are functions of the free observables, determined by the parametrizations of two-loop radiative corrections in Refs.~\cite{Awramik:2003rn,Awramik:2006uz,Freitas:2014hra}.
The new physics contributions are computed with the
oblique parameters~\cite{Ciuchini:2013pca}:
\begin{eqnarray}
\Delta {m_W} &=&  - \frac{{\alpha m_W^{{\mathrm{SM}}}}}{{4(c_{\mathrm{W}}^2 - s_{\mathrm{W}}^2)}}{F_1}(S,T,U),
\\
\Delta {\sin ^2}\theta _{{\mathrm{eff}}}^\ell  &=& \frac{\alpha }{{4(c_{\mathrm{W}}^2 - s_{\mathrm{W}}^2)}}{F_2}(S,T),
\\
\Delta {\Gamma _Z} &=& \frac{{{\alpha ^2}{m_Z}}}{{72s_{\mathrm{W}}^2c_{\mathrm{W}}^2(c_{\mathrm{W}}^2 - s_{\mathrm{W}}^2)}}{F_3}(S,T).
\end{eqnarray}

\begin{table}[!t]
\centering
\setlength\tabcolsep{.4em}
\renewcommand{\arraystretch}{1.3}
\begin{tabular}{|c|cc|}
\hline
 & Current data & CEPC-B precision \\
\hline
$\alpha_\mathrm{s}(m_Z^2)$ & $0.1185 \pm 0.0006$~\cite{Agashe:2014kda} & $\pm 1 \times 10^{-4}$~\cite{Lepage:2014fla} \\
$\Delta \alpha _{{\mathrm{had}}}^{(5)}(m_Z^2)$ & $0.02765 \pm 0.00008$~\cite{Bodenstein:2012pw} & $ \pm 4.7 \times {10^{ - 5}}$~\cite{Jegerlehner:2003ip,Baak:2014ora} \\
$m_Z$ [GeV] & $91.1875 \pm 0.0021$~\cite{ALEPH:2005ab} & $ \pm 5 \times 10^{-4}$~\cite{CEPC-SPPCStudyGroup:2015csa} \\
$m_t$ [GeV] & $173.34 \pm 0.76_\mathrm{ex}$~\cite{ATLAS:2014wva}~$\pm 0.5_\mathrm{th}$~\cite{Erler:2014eya} & $ \pm 0.2_\mathrm{ex}$~\cite{CMS:2013wfa}~$\pm 0.5_\mathrm{th}$~\cite{Erler:2014eya} \\
$m_h$ [GeV] & $125.09 \pm 0.24$~\cite{Aad:2015zhl} & $ \pm 5.9 \times 10^{-3}$~\cite{CEPC-SPPCStudyGroup:2015csa} \\
$m_W$ [GeV] & $80.385 \pm 0.015_\mathrm{ex}$~\cite{Agashe:2014kda}~$ \pm 0.004_\mathrm{th}$~\cite{Awramik:2003rn} & $(\pm 3_\mathrm{ex}$~\cite{CEPC-SPPCStudyGroup:2015csa}~$\pm 1_\mathrm{th}$~\cite{Freitas:2013xga}$)\times 10^{-3}$ \\
${\sin ^2}\theta _{{\mathrm{eff}}}^\ell$ & $0.23153 \pm 0.00016$~\cite{ALEPH:2005ab} & $(\pm 2.3_\mathrm{ex}$~\cite{CEPC-SPPCStudyGroup:2015csa}~$\pm 1.5_\mathrm{th}$~\cite{Freitas:2013xga}$)\times 10^{-5}$ \\
$\Gamma_Z$ [GeV] & $2.4952 \pm 0.0023$~\cite{ALEPH:2005ab} & $(\pm 5_\mathrm{ex}$~\cite{CEPC-SPPCStudyGroup:2015csa}~$\pm 0.8_\mathrm{th}$~\cite{Fan:2014vta,Mishima}$)\times 10^{-4}$  \\
\hline
\end{tabular}
\caption{Current measurement values and CEPC baseline (CEPC-B) precisions of EW precision observables. The subscripts ``ex'' and ``th'' denote experimental and theoretical uncertainties, respectively. For the unspecified uncertainties, theoretical uncertainties are either neglected or incorporated.}
\label{tab:CEPC-B}
\end{table}

Table~\ref{tab:CEPC-B} shows the current measurement values and CEPC baseline precisions (denoted as ``CEPC-B'' hereafter) of the eight EW precision observables. The references for these values are also listed in the table. Experimental (``ex'') and theoretical (``th'') uncertainties are separately denoted for some values. For those unspecified uncertainties, theoretical uncertainties are either neglected or incorporated into the total uncertainties.

Experimental uncertainties for the CEPC-B precisions will be mostly reduced by the running of CEPC, according to the preliminary conceptual design report of CEPC~\cite{CEPC-SPPCStudyGroup:2015csa}. Exceptions are the potential reduction of the $\alpha_\mathrm{s}(m_Z^2)$ uncertainty due to lattice QCD calculation in the next decade~\cite{Lepage:2014fla},  the improvement of the $\Delta \alpha _{{\mathrm{had}}}^{(5)}(m_Z^2)$ measurement from ongoing charm and bottom factories as well as lattice QCD prediction~\cite{Baak:2014ora}, and the improvement of the $m_t$ measurement at the high-luminosity LHC~\cite{CMS:2013wfa}.
We also consider that the theoretical uncertainties of $m_W$, ${\sin ^2}\theta _{{\mathrm{eff}}}^\ell$, and $\Gamma_Z$ can be reduced by fully calculating three-loop corrections in the future~\cite{Freitas:2013xga,Fan:2014vta,Mishima}.

\begin{table}[!t]
\centering
\setlength\tabcolsep{.4em}
\renewcommand{\arraystretch}{1.3}
\begin{tabular}{|c|c|}
\hline
 & CEPC-I precision \\
\hline
$m_Z$ [GeV] & $ \pm 1 \times 10^{-4}$~\cite{CEPC-SPPCStudyGroup:2015csa} \\
$\Gamma_Z$ [GeV] & $(\pm 1_\mathrm{ex}$~\cite{CEPC-SPPCStudyGroup:2015csa}~$\pm 0.8_\mathrm{th}$~\cite{Fan:2014vta,Mishima}$)\times 10^{-4}$ \\
$m_t$ [GeV] & $ \pm 0.03_\mathrm{ex} \pm 0.1_\mathrm{th}$~\cite{Baer:2013cma} \\
\hline
\end{tabular}
\caption{CEPC improved (CEPC-I) precisions of $m_Z$, $\Gamma_Z$, and $m_t$ taking into account their potential improvements.}
\label{tab:CEPC-I}
\end{table}

There may be some further improvements. A high-precision calibration of the beam energy may be achieved at the CEPC, reducing the experimental uncertainties of $m_Z$ and $\Gamma_Z$ down to $0.1~\si{MeV}$~\cite{CEPC-SPPCStudyGroup:2015csa}.
The current CEPC plan does not include a $t\bar{t}$ threshold scan.
Nevertheless, if the ILC would preform this kind of scan before or during the running of CEPC, the experimental and theoretical uncertainties of $m_t$ could be reduced to $30$ and $100~\si{MeV}$, respectively~\cite{Baer:2013cma}.
These potential improvements are collected in Table~\ref{tab:CEPC-I}, and the corresponding precisions are denoted as ``CEPC-I'' hereafter.

\subsection{Global Fits}

We calculate a modified $\chi^2$ function for global fits~\cite{Fan:2014vta}:
\begin{eqnarray}\label{eq:chi2}
\chi _{{\mathrm{mod}}}^2 &=& \sum\limits_{i} {{{\left( {\frac{{O_i^{{\mathrm{meas}}} - O_i^{{\mathrm{pred}}}}}{{{\sigma _i}}}} \right)}^2}}
\nonumber\\
&&  + \sum\limits_{j} {\left\{ { - 2\ln \left[ {{\mathrm{erf}}\left( {\frac{{O_j^{{\mathrm{meas}}} - O_j^{{\mathrm{pred}}} + {\delta _j}}}{{\sqrt 2 {\sigma _j}}}} \right) - {\mathrm{erf}}\left( {\frac{{O_j^{{\mathrm{meas}}} - O_j^{{\mathrm{pred}}} - {\delta _j}}}{{\sqrt 2 {\sigma _j}}}} \right)} \right]} \right\}}, \qquad
\end{eqnarray}
where $O_i^\mathrm{meas}$ and $O_i^\mathrm{pred}$ denote the measured and predicted values of the observables. For the free observables, we take the mean values of the current measurements as $O_i^\mathrm{meas}$. Meanwhile, $O_i^\mathrm{meas}$ for the induced observables $m_W$, ${\sin ^2}\theta _{{\mathrm{eff}}}^\ell$, and $\Gamma_Z$ are set to be their SM prediction values. By this way, the mean values of the oblique parameters in the fit result will locate at zero, and thus the current and CEPC precisions are just represented by the uncertainties of $S$, $T$, and $U$, which will be expressed by standard deviations and correlation coefficients.
The first term in Eq.~\eqref{eq:chi2} is an ordinary $\chi^2$, corresponding to the observables whose uncertainties $\sigma_i$ are not split into two parts in Tables~\ref{tab:CEPC-B} and \ref{tab:CEPC-I}. The other observables belong to the second term, which treats the experimental uncertainties $\sigma_j$ as Gaussian errors while the theoretical uncertainties $\delta_j$ as flat box-shaped errors, following Refs.~\cite{Hocker:2001xe,Flacher:2008zq,Lafaye:2009vr,Fan:2014vta}.

\begin{table}[!t]
\centering
\setlength\tabcolsep{.4em}
\renewcommand{\arraystretch}{1.3}
\begin{tabular}{|c|ccc|ccc|}
\hline
 & $\sigma_S$ & $\sigma_T$ & $\sigma_U$ & $\rho_{ST}$ & $\rho_{SU}$ & $\rho_{TU}$ \\
\hline
Current & 0.10  & 0.12   & 0.094 & $+0.89$ & $-0.55$ & $-0.80$ \\
CEPC-B  & 0.021 & 0.026  & 0.020 & $+0.90$ & $-0.68$ & $-0.84$ \\
CEPC-I  & 0.011 & 0.0071 & 0.010 & $+0.74$ & $+0.15$ & $-0.21$ \\
\hline
\end{tabular}
\caption{Fit results for $S$, $T$, and $U$ with current, CEPC-B, and CEPC-I precisions. $\sigma_i$ and $\rho_{ij}$ ($i,j=S,T,U$) are the standard deviations and the correlation coefficients, respectively.}
\label{tab:STU}
\end{table}

\begin{figure}[!t]
\centering
\includegraphics[width=0.7\textwidth]{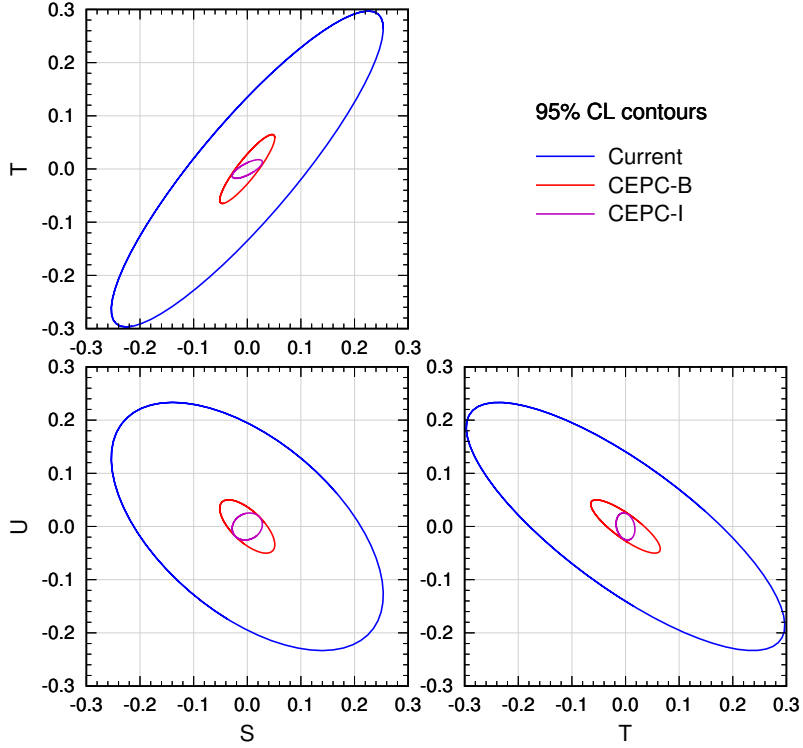}
\caption{95\% CL contours in the $S-T$, $S-U$, and $T-U$ planes for current, CEPC-B, and CEPC-I precisions.}
\label{fig:cont:STU}
\end{figure}

We utilize the code \texttt{MultiNest}~\cite{Feroz:2013hea} to perform a quick and stable global fit.
Firstly, we treat all of $S$, $T$, and $U$ as free parameters, and obtain the fit results presented in Table~\ref{tab:STU}. Fig.~\ref{fig:cont:STU} demonstrates the corresponding 95\% CL contours in the $S-T$, $S-U$, and $T-U$ planes.
We can see that the running of CEPC will greatly improve the precision of the oblique parameters.
Correlation relations among these parameters are not quite definite: the sign of $\rho_{SU}$ in the CEPC-I precisions is different from those in the current and CEPC-B precisions.
Nonetheless, the correlation between $S$ and $T$ seems positive and close to one.
This can be easily understood from Eqs.~\eqref{eq:F1}--\eqref{eq:F2}, whose numerical results are
\begin{equation}\label{eq:F_i:num}
F_1=S-1.55T-1.24U, \quad F_2=S-0.69T, \quad F_3=-12.2(S-2.7T).
\end{equation}
Therefore, the increase of $S$ can be always compensated by increasing $T$, leading to a high positive correlation~\cite{Erler:1994fz,Agashe:2014kda}.

\begin{table}[!t]
\centering
\setlength\tabcolsep{.4em}
\renewcommand{\arraystretch}{1.3}
\subtable[~$U=0$ fixed]{
\begin{tabular}{|c|cc|c|}
\hline
 & $\sigma_S$ & $\sigma_T$ & $\rho_{ST}$ \\
\hline
Current & 0.085 & 0.072  & $+0.90$ \\
CEPC-B  & 0.015 & 0.014  & $+0.83$ \\
CEPC-I  & 0.011 & 0.0069 & $+0.80$ \\
\hline
\end{tabular}
}
\hspace{2em}
\subtable[~$S=0$ fixed]{
\begin{tabular}{|c|cc|c|}
\hline
 & $\sigma_T$ & $\sigma_U$ & $\rho_{TU}$ \\
\hline
Current & 0.054  & 0.078  & $-0.81$ \\
CEPC-B  & 0.011  & 0.015  & $-0.72$ \\
CEPC-I  & 0.0048 & 0.010  & $-0.48$ \\
\hline
\end{tabular}
}
\tabcaption{Fit results with current, CEPC-B, and CEPC-I precisions under the assumptions of $U=0$ (a) and $S=0$ (b).
$\sigma_i$ and $\rho_{ij}$ ($i,j=S,T,U$) are the standard deviations and the correlation coefficients, respectively.}
\label{tab:ST:TU}
\end{table}

\begin{figure}[!t]
\centering
\subfigure[~$U=0$ fixed]
{\includegraphics[width=0.49\textwidth]{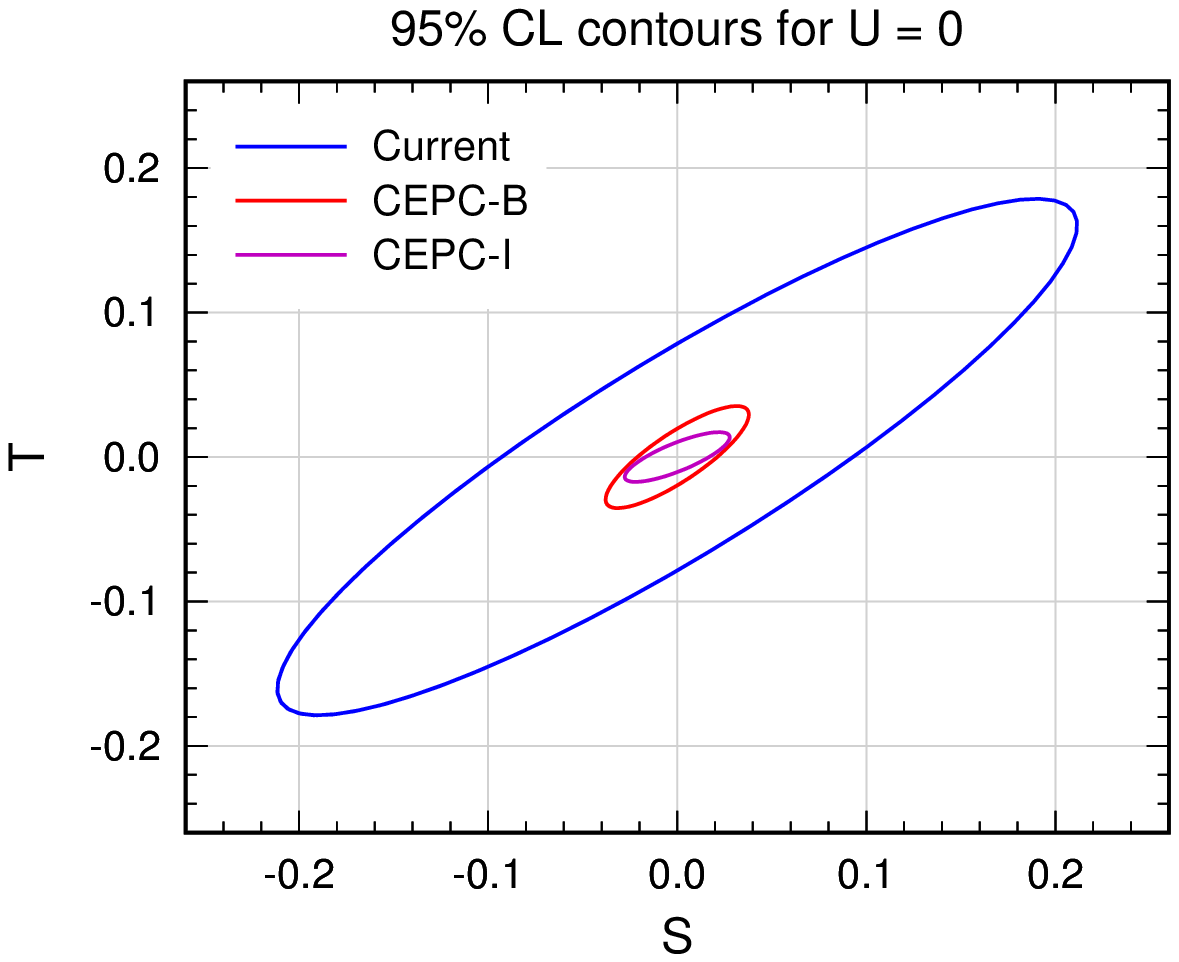}}
\subfigure[~$S=0$ fixed]
{\includegraphics[width=0.49\textwidth]{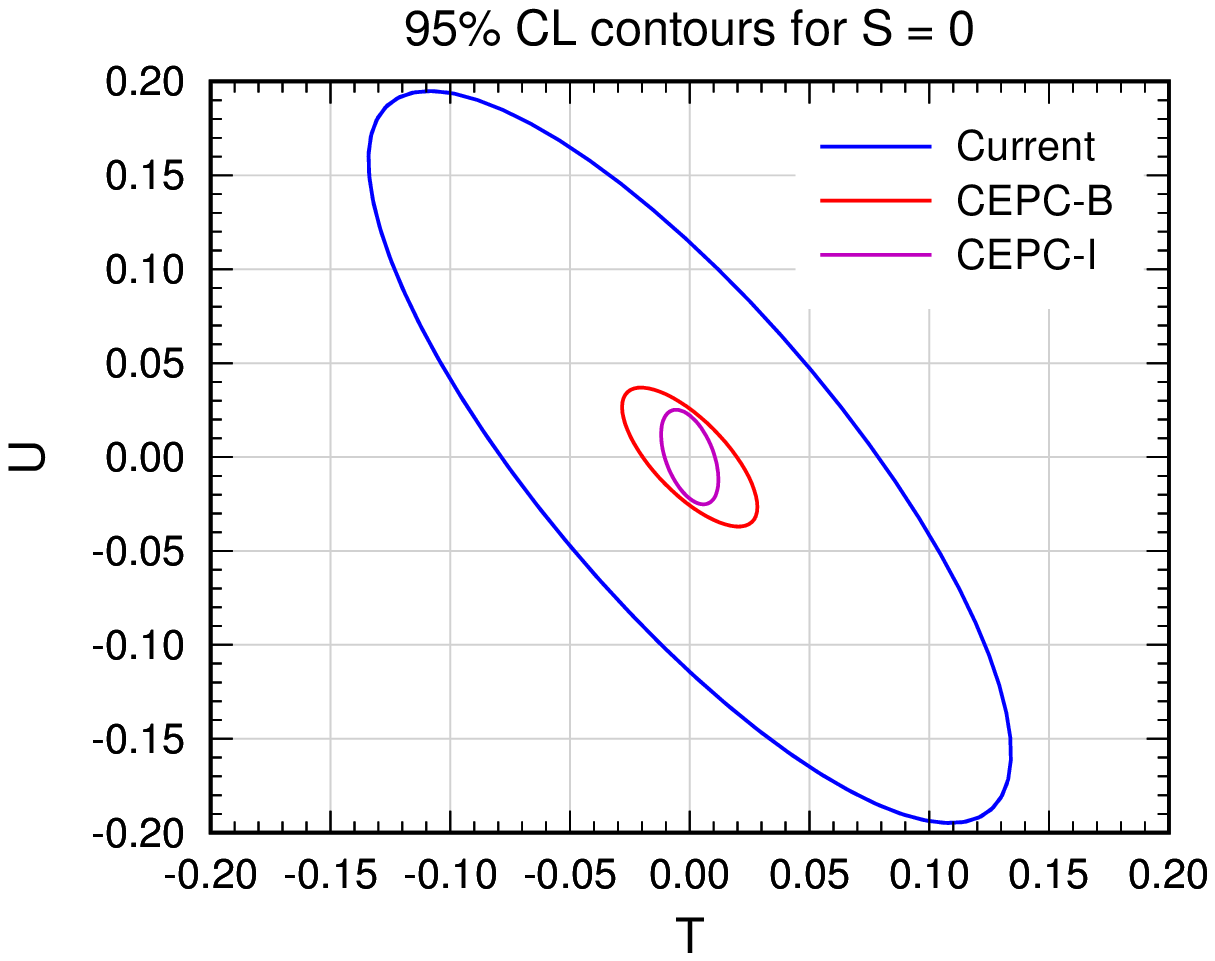}}
\figcaption{95\% CL contours in the $S-T$ (a) and $T-U$ (b) plane for current, CEPC-B, and CEPC-I precisions under the assumptions of $U=0$ (a) and $S=0$ (b).}
\label{fig:cont:ST:TU}
\end{figure}

Moreover, we carry out the global fit with some of the oblique parameters fixed to zero.
We separately consider two assumptions by fixing one parameter to zero: (a) the assumption of $U=0$, which is useful for new physics models predicting a tiny $U$;
(b) the assumption of $S=0$, which often holds for introducing new $\mathrm{SU}(2)_\mathrm{L}$ multiplets with zero hypercharge.
The fit results are listed in Table~\ref{tab:ST:TU} and the corresponding contours in the $S-T$ or $T-U$ plane at 95\% CL are shown in Fig.~\ref{fig:cont:ST:TU}.
For the $U=0$ case, $S$ and $T$ always have a high positive correlation as expected.
The result is consistent with those given in Refs.~\cite{Fan:2014vta,CEPC-SPPCStudyGroup:2015csa}.
For the $S=0$ case, the $F_1$ function in Eq.~\eqref{eq:F_i:num} results in a negative correlation between $T$ and $U$.

\begin{table}[!t]
\centering
\setlength\tabcolsep{.4em}
\renewcommand{\arraystretch}{1.3}
\subtable[~$T=U=0$ fixed]{
\begin{tabular}{|c|c|}
\hline
 & $\sigma_S$ \\
\hline
Current & 0.037 \\
CEPC-B  & 0.0085 \\
CEPC-I  & 0.0068 \\
\hline
\end{tabular}
}
\hspace{6em}
\subtable[~$S=U=0$ fixed]{
\begin{tabular}{|c|c|}
\hline
 & $\sigma_T$ \\
\hline
Current & 0.032  \\
CEPC-B  & 0.0079  \\
CEPC-I  & 0.0042 \\
\hline
\end{tabular}
}
\tabcaption{Fit results with current, CEPC-B, and CEPC-I precisions under the assumptions of $T=U=0$ (a) and $S=U=0$ (b).
$\sigma_S$ and $\sigma_T$ are the standard deviations of $S$ and $T$, respectively.}
\label{tab:S:T}
\end{table}

We also present fit results for fixing two oblique parameters to zero.
In Table~\ref{tab:S:T}(a), the results for the $S$ parameter are obtained under the assumption of $T=U=0$, which corresponds to the models that respect a custodial symmetry.
In Table~\ref{tab:S:T}(b), we give the results for the $T$ parameter under the assumption of $S=U=0$. They are useful for the models that contain new $\mathrm{SU}(2)_\mathrm{L}$ multiplets with zero hypercharge and also predict $U\ll T$.

Below, we use the above results to estimate the expected constraints on the fermionic WIMP models.

\section{Singlet-Doublet Fermionic Dark Matter}
\label{sec:SDFDM}

\subsection{Fields and Interactions}

In the SDFDM model, we introduce three left-handed Weyl spinors:~\cite{Mahbubani:2005pt,D'Eramo:2007ga,Enberg:2007rp,Cohen:2011ec,Calibbi:2015nha}
\begin{equation}
S\in (\textbf{1},0), \quad D_1=\begin{pmatrix}D_1^0\\D_1^-\end{pmatrix}\in(\textbf{2},-1/2),\quad D_2=\begin{pmatrix}D_2^+\\D_2^0\end{pmatrix}\in(\textbf{2},+1/2).
\end{equation}
Their gauge transformations under $(\mathrm{SU}(2)_\mathrm{L},\mathrm{U}(1)_\mathrm{Y})$ are denoted. The kinetic and interacting properties are encoded in the following Lagrangians:
\begin{eqnarray}
\mathcal{L}_\mathrm{S}&=&iS^\dag\bar{\sigma}^\mu\partial_\mu S-\frac{1}{2}(m_SSS+\mathrm{h.c.}),\\
\mathcal{L}_\mathrm{D}&=&iD_{1}^\dag\bar{\sigma}^\mu D_\mu D_1+iD_{2}^\dag\bar{\sigma}^\mu D_\mu D_2-(m_D\epsilon_{ij}D_1^iD_2^j+\mathrm{h.c.}),\\
\mathcal{L}_\mathrm{HSD}&=&y_1H_iSD_1^i-y_2H_i^\dag S D_2^i+\mathrm{h.c.}\,,
\end{eqnarray}
where $H$ is the SM Higgs doublet and $D_\mu$ is the covariant derivative.
Gauge interactions of the doublets are
\begin{eqnarray}\label{LD}
\mathcal{L}_\mathrm{D}&\supset&\frac{g}{\sqrt{2}}[W_\mu^+(D_1^0)^\dag\bar{\sigma}^\mu D_1^-+W_\mu^-(D_1^-)^\dag\bar{\sigma}^\mu D_1^0
+W_\mu^+(D_2^+)^\dag\bar{\sigma}^\mu D_2^0+W_\mu^-(D_2^0)^\dag\bar{\sigma}^\mu D_2^+]\nonumber\\
&&+\frac{g}{2c_\mathrm{W}}Z_\mu(D_1^0)^\dag\bar{\sigma}^\mu D_1^0+\left[-eA_\mu+\frac{g}{2c_\mathrm{W}}(s_\mathrm{W}^2-c_\mathrm{W}^2)Z_\mu\right](D_1^-)^\dag\bar{\sigma}^\mu D_1^-\nonumber\\
&&-\frac{g}{2c_\mathrm{W}}Z_\mu(D_2^0)^\dag\bar{\sigma}^\mu D_2^0+\left[eA_\mu-\frac{g}{2c_\mathrm{W}}(s_\mathrm{W}^2-c_\mathrm{W}^2)Z_\mu\right](D_2^+)^\dag\bar{\sigma}^\mu D_2^+.
\end{eqnarray}

In the unitary gauge, $H=(0,(v+h)/\sqrt{2})^\mathrm{T}$ with the VEV $v\approx 246~\si{GeV}$, and the mass terms are
\begin{eqnarray}
\mathcal{L}_\mathrm{mass}&=&-\frac{1}{2}(S\quad D_1^0\quad D_2^0)\mathcal{M}_\mathrm{N}\begin{pmatrix}S\\D_1^0\\D_2^0\end{pmatrix}-m_DD_1^-D_2^++\mathrm{h.c.}\nonumber\\
&=&-\frac{1}{2}\sum_{i=1}^3m_{\chi_i^0}\chi_i^0\chi_i^0-m_{\chi^\pm}\chi^-\chi^++\mathrm{h.c.}\,,
\end{eqnarray}
where we define the mass matrix and the fields as
\begin{eqnarray}
&&\mathcal{M}_\mathrm{N}=\begin{pmatrix}
m_S&\dfrac{1}{\sqrt{2}}y_1v&\dfrac{1}{\sqrt{2}}y_2v\\[1em] \dfrac{1}{\sqrt{2}}y_1v&0&-m_D\\[1em] \dfrac{1}{\sqrt{2}}y_2v&-m_D&0\end{pmatrix},\quad
m_{\chi^\pm}=m_D,\quad
\chi^+=D_2^+,\quad
\chi^-=D_1^-,\\
&&\mathcal{N}^\mathrm{T}\mathcal{M}_\mathrm{N}\mathcal{N}=\mathrm{diag}(m_{\chi_1^0},m_{\chi_2^0},m_{\chi_3^0}),\quad
\begin{pmatrix}S\\D_1^0\\D_2^0\end{pmatrix}=\mathcal{N}\begin{pmatrix}\chi_1^0\\ \chi_2^0\\ \chi_3^0\end{pmatrix}.
\end{eqnarray}
Thus, the new mass states are one singly charged fermion $\chi^\pm$ and three Majorana fermions $\chi_{1,2,3}^0$, where the lightest neutral fermion $\chi_1^0$ serves as a DM candidate.

The Lagrangian for the trilinear interaction between $\chi_1^0$ and the Higgs boson is
\begin{equation}\label{eq:L_hchichi}
\mathcal{L}_{h\chi_1^0\chi_1^0}=\frac{1}{2}g_{h\chi_1^0\chi_1^0}h\bar\chi_1^0\chi_1^0,
\end{equation}
where the $h\chi_1^0\chi_1^0$ coupling is given by
\begin{equation}\label{gh00}
g_{h\chi_1^0\chi_1^0}=-\sqrt{2}(y_1\mathcal{N}_{21}+y_2\mathcal{N}_{31})\mathcal{N}_{11}.
\end{equation}
This coupling induces spin-independent (SI) DM-nucleus scattering.
Since $\chi_1^0$ is a Majorana fermion, the vector current operator $\bar\chi_1^0\gamma^\mu\chi_1^0$ vanishes. Thus, $\chi_1^0$ can only couple to $Z$ through an axial current interaction Lagrangian
\begin{equation}\label{eq:L_Zchichi}
\mathcal{L}_{Z\chi_1^0\chi_1^0}=\frac{1}{2}g_{Z\chi_1^0\chi_1^0}Z_\mu\bar\chi_1^0\gamma^\mu\gamma_5\chi_1^0,
\end{equation}
where the $Z\chi_1^0\chi_1^0$ coupling is
\begin{equation}\label{gZ00}
g_{Z\chi_1^0\chi_1^0}=-\frac{g}{2c_\mathrm{W}}(|\mathcal{N}_{21}|^2-|\mathcal{N}_{31}|^2).
\end{equation}
This coupling will not induce SI scattering, but it leads to spin-dependent (SD) scattering.
Direct detection experiments search for recoil signals from DM-nucleus scattering and could be sensitive to $\chi_1^0$. Related formulas are collected in Appendix~\ref{app:DM_DD}.

\subsection{Vacuum Polarizations and Custodial Symmetry}

The dark sector fermions affect the vacuum polarizations of EW gauge bosons at one-loop level, and hence contribute to the EW oblique parameters $S$, $T$, and $U$.
Their contributions to the vacuum polarizations are given by
\begin{eqnarray}
\Pi_{AA}(p^2)&=&\frac{2e^2}{16\pi^2}J_2(p^2,m_{\chi^\pm}^2),\qquad
\Pi_{ZA}(p^2)=\frac{2eg_{Z\chi^+\chi^-}}{16\pi^2}J_2(p^2,m_{\chi^\pm}^2),\\
\Pi_{ZZ}(p^2)&=&\frac{1}{16\pi^2}\bigg\{\sum_{i,j=1}^{3}\big[|g_{Z\chi_i^0\chi_j^0}|^2J_1(p^2,m_{\chi_i^0}^2,m_{\chi_j^0}^2)\nonumber\\
&&\qquad\qquad\qquad+m_{\chi_i^0}m_{\chi_j^0}(g_{Z\chi_i^0\chi_j^0}^2+\mathrm{h.c.})B_0(p^2,m_{\chi_i^0}^2,m_{\chi_j^0}^2)\big]\nonumber\\
&&\qquad\quad+2g_{Z\chi^+\chi^-}^2\big[J_1(p^2,m_{\chi^\pm}^2,m_{\chi^\pm}^2)-2m_{\chi^\pm}^2B_0(p^2,m_{\chi^\pm}^2,m_{\chi^\pm}^2)\big]\bigg\},\quad\\
\Pi_{WW}(p^2)&=&\frac{1}{16\pi^2}\sum_{j=1}^3\big[(|a_{W\chi^+\chi_j^0}|^2+|b_{W\chi^+\chi_j^0}|^2)J_1(p^2,m_{\chi_j^0}^2,m_{\chi^\pm}^2)\nonumber\\
&&\qquad\qquad\quad-2m_{\chi_j^0}m_{\chi^\pm}(a_{W\chi^+\chi_j^0}b_{W\chi^+\chi_j^0}^\ast+\mathrm{h.c.})B_0(p^2,m_{\chi_j^0}^2,m_{\chi^\pm}^2)\big].
\end{eqnarray}
We define couplings
\begin{eqnarray}
g_{Z\chi^+\chi^-}&=&\frac{g(c_\mathrm{W}^2-s_\mathrm{W}^2)}{2c_\mathrm{W}},\quad
g_{Z\chi^0_i\chi^0_j}=-\frac{g}{2c_\mathrm{W}}(\mathcal{N}_{2,i}\mathcal{N}_{2,j}^\ast-\mathcal{N}_{3,i}\mathcal{N}_{3,j}^\ast),\\
a_{W\chi^+\chi_j^0}&=&\frac{g}{\sqrt{2}}\mathcal{N}_{3,j},\quad
b_{W\chi^+\chi_j^0}=-\frac{g}{\sqrt{2}}\mathcal{N}_{2,j}^\ast,
\end{eqnarray}
and functions
\begin{eqnarray}\label{J1J2}
J_1(p^2,m_1^2,m_2^2)&=&A_0(m_1^2)+A_0(m_2^2)-(p^2-m_1^2-m_2^2)B_0(p^2,m_1^2,m_2^2)\nonumber\\
&&-4B_{00}(p^2,m_1^2,m_2^2),\\
J_2(p^2,m^2)&=&J_1(p^2,m^2,m^2)-2m^2B_0(p^2,m^2,m^2)\nonumber\\
&=&2A_0(m^2)-p^2B_0(p^2,m^2,m^2)-4B_{00}(p^2,m^2,m^2),
\end{eqnarray}
where the Passiano-Veltman scalar functions~\cite{Passarino:1978jh} have consistent definitions with Ref.~\cite{Denner:1991kt}:
\begin{eqnarray}
{A_0}({m^2}) &=& \frac{{{{(2\pi Q)}^{4 - d}}}}{{i{\pi ^2}}}\int {{d^d}q} \frac{1}{{{q^2} - {m^2} + i\varepsilon }},
\\
{B_0}({p^2},m_1^2,m_2^2) &=& \frac{{{{(2\pi Q)}^{4 - d}}}}{{i{\pi ^2}}}\int {{d^d}q} \frac{1}{{[{q^2} - m_1^2 + i\varepsilon ][{{(q + p)}^2} - m_2^2 + i\varepsilon ]}},
\\
{g_{\mu \nu }}{B_{00}}({p^2},m_1^2,m_2^2) + {p_\mu }{p_\nu }{B_{11}}({p^2},m_1^2,m_2^2) \hspace*{-9em} &&
\nonumber\\
&=& \frac{{{{(2\pi Q)}^{4 - d}}}}{{i{\pi ^2}}}\int {{d^d}q} \frac{{{q_\mu }{q_\nu }}}{{[{q^2} - m_1^2 + i\varepsilon ][{{(q + p)}^2} - m_2^2 + i\varepsilon ]}}.\qquad
\end{eqnarray}
We use \texttt{LoopTools}~\cite{Hahn:1998yk} to give numerical values for these functions.

For $m_1=m_2=m$, we have
\begin{eqnarray}
J'_1(0,m^2,m^2)&=&-\frac{2}{3}\Delta+\frac{2}{3}\ln\frac{m^2}{Q^2}+\frac{1}{3},\label{eq:equalmass:J1}\\
B'_0(0,m^2,m^2)&=&\frac{1}{6 m^2},\label{eq:equalmass:B0}
\end{eqnarray}
where $\Delta\equiv 2/(4-d)-\gamma_\mathrm{E}+\ln 4\pi$ is the UV-divergent term.
If $m_1\ll m_2$, the following approximations hold:
\begin{eqnarray}
J'_1(0,m_1^2,m_2^2)&\approx&-\frac{2}{3}\Delta+\frac{2}{3}\ln\frac{m_2^2}{Q^2}-\frac{2}{9}+\mathcal{O}\left(\frac{m_1^2}{m_2^2}\right),\label{eq:appr:J1}\\
B'_0(0,m_1^2,m_2^2)&\approx&\frac{1}{2m_2^2}+\mathcal{O}\left(\frac{m_1^2}{m_2^4}\right).\label{eq:appr:B0}
\end{eqnarray}
These expressions are useful for the analyses below.

When $y_1=y_2=y$, there is a custodial global symmetry in this model. It can be clarified by defining $\mathrm{SU}(2)_\mathrm{R}$ doublets
\begin{equation}
(\mathcal{D}^A)^i=\begin{pmatrix}D_1^i\\ D_2^i\end{pmatrix},\quad
(\mathcal{H}^A)_i=\begin{pmatrix}H_i^\dag\\ H_i\end{pmatrix},
\end{equation}
since the Lagrangians has $\mathrm{SU}(2)_\mathrm{L}\times\mathrm{SU}(2)_\mathrm{R}$ invariant forms
\begin{eqnarray}
\mathcal{L}_\mathrm{D}
&=&i\mathcal{D}_A^\dag\bar{\sigma}^\mu D_\mu\mathcal{D}^A-\frac{1}{2}[m_D\epsilon_{AB}\epsilon_{ij}(\mathcal{D}^A)^i(\mathcal{D}^B)^j+\mathrm{h.c.}],\\
\mathcal{L}_\mathrm{HSD}&=&y(H_iSD_1^i-H_i^\dag SD_2^i)+\mathrm{h.c.}=y\epsilon_{AB}(\mathcal{H}^A)_iS(\mathcal{D}^B)^j+\mathrm{h.c.}\,.
\end{eqnarray}
Therefore, it is expected to have vanishing $T$ and $U$ in this custodial symmetry limit.

Moreover, there are other important implications in this limit: at tree level, the SD DM-nucleon scattering cross section vanishes, and the SI scattering cross section vanishes as well if $m_S>m_D$.
The first implication can be easily understood. The custodial symmetry ensures the up and down components of the $\mathrm{SU}(2)_\mathrm{R}$ doublet $\mathcal{D}_A$ have equal Dirac mass terms induced by the nonzero VEV. Consequently, each neutral mass state $\chi_i^0$ has equal $D_1$ and $D_2$ components ($|\mathcal{N}_{21}|=|\mathcal{N}_{31}|$). Since $D_1$ and $D_2$ have opposite hypercharges and opposite third components of weak isospin, the $Z\chi_i^0\chi_i^0$ couplings becomes zero due to the exact cancellation, leading to a vanishing SD scattering cross section.
The second implication is not as obvious as the first one, but we can understand both through the following analysis.

In the $y_1=y_2=y$ limit, if $m_S>m_D$, the mass matrix $\mathcal{M}_\mathrm{N}$ can be diagonalized by the mixing matrix
\begin{equation}\label{NSD}
\mathcal{N}=\begin{pmatrix}0&is_\beta&c_\beta\\[.2em] \dfrac{1}{\sqrt{2}}&-\dfrac{ic_\beta}{\sqrt{2}}&\dfrac{s_\beta}{\sqrt{2}}\\[1em] -\dfrac{1}{\sqrt{2}}&-\dfrac{ic_\beta}{\sqrt{2}}&\dfrac{s_\beta}{\sqrt{2}}\end{pmatrix},
\end{equation}
where $s_\beta$ and $c_\beta$ are some real numbers satisfying $s_\beta^2+c_\beta^2=1$.
The mass eigenvalues are
\begin{eqnarray}
m_{\chi_1^0}&=&m_D,\\
m_{\chi_2^0}&=&\frac{1}{2}[\sqrt{(m_D+m_S)^2+4y^2v^2}-m_S+m_D],\\
m_{\chi_3^0}&=&\frac{1}{2}[\sqrt{(m_D+m_S)^2+4y^2v^2}+m_S-m_D].
\end{eqnarray}
Substituting $\mathcal{N}_{11}=0$ and  $\mathcal{N}_{21}=-\mathcal{N}_{31}=1/\sqrt{2}$ into Eqs.~\eqref{gh00} and \eqref{gZ00}, one finds $g_{Z\chi_1^0\chi_1^0}=g_{h\chi_1^0\chi_1^0}=0$.
Therefore, direct detection experiments can hardly constrain the model in this case.

If $m_S<m_D$, the mass eigenvalues become
\begin{eqnarray}
m_{\chi_{1}^0}&=&\min\left(m_D,\,\frac{1}{2}[\sqrt{(m_D+m_S)^2+4y^2v^2}+m_S-m_D]\right),\\
m_{\chi_{2}^0}&=&\max\left(m_D,\,\frac{1}{2}[\sqrt{(m_D+m_S)^2+4y^2v^2}+m_S-m_D]\right),\\
m_{\chi_3^0}&=&\frac{1}{2}[\sqrt{(m_D+m_S)^2+4y^2v^2}-m_S+m_D].
\end{eqnarray}
There are two kinds of mass order.
If $m_{\chi_{1}^0}=m_D$, the $\mathcal{N}$ matrix remains the form of Eq.~\eqref{NSD}, leading to $g_{Z\chi_1^0\chi_1^0}=g_{h\chi_1^0\chi_1^0}=0$.
Otherwise we should multiply \eqref{NSD} by a permutation matrix to obtain the correct mass order:
\begin{equation}
\mathcal{N}\to\mathcal{N}\begin{pmatrix}0&1&0\\ 1&0&0\\0&0&1\end{pmatrix}
=\begin{pmatrix}is_\beta&0&c_\beta\\[.2em] -\dfrac{ic_\beta}{\sqrt{2}}&\dfrac{1}{\sqrt{2}}&\dfrac{s_\beta}{\sqrt{2}}\\[1em] -\dfrac{ic_\beta}{\sqrt{2}}&-\dfrac{1}{\sqrt{2}}&\dfrac{s_\beta}{\sqrt{2}}\end{pmatrix}.
\end{equation}
This still leads to a vanishing $g_{Z\chi_1^0\chi_1^0}$, but $g_{h\chi_1^0\chi_1^0}$ becomes nonzero.
Thus, there will be some constraints from SI direct detection.

Furthermore, $y_1=-y_2$ corresponds to another custodial symmetry limit, which can easily examined by instead defining $(\mathcal{H}^A)_i=(-H_i^\dag, H_i)$.
In this limit, we also have $T=U=0$ and $g_{Z\chi_1^0\chi_1^0}=0$.

\subsection{Expected Constraints}
\label{subsec:SDFDM:cstr}

\begin{figure}[!t]
\centering
\subfigure[~$m_S<m_D$ case\label{SDSTU:a}]
{\includegraphics[width=0.49\textwidth]{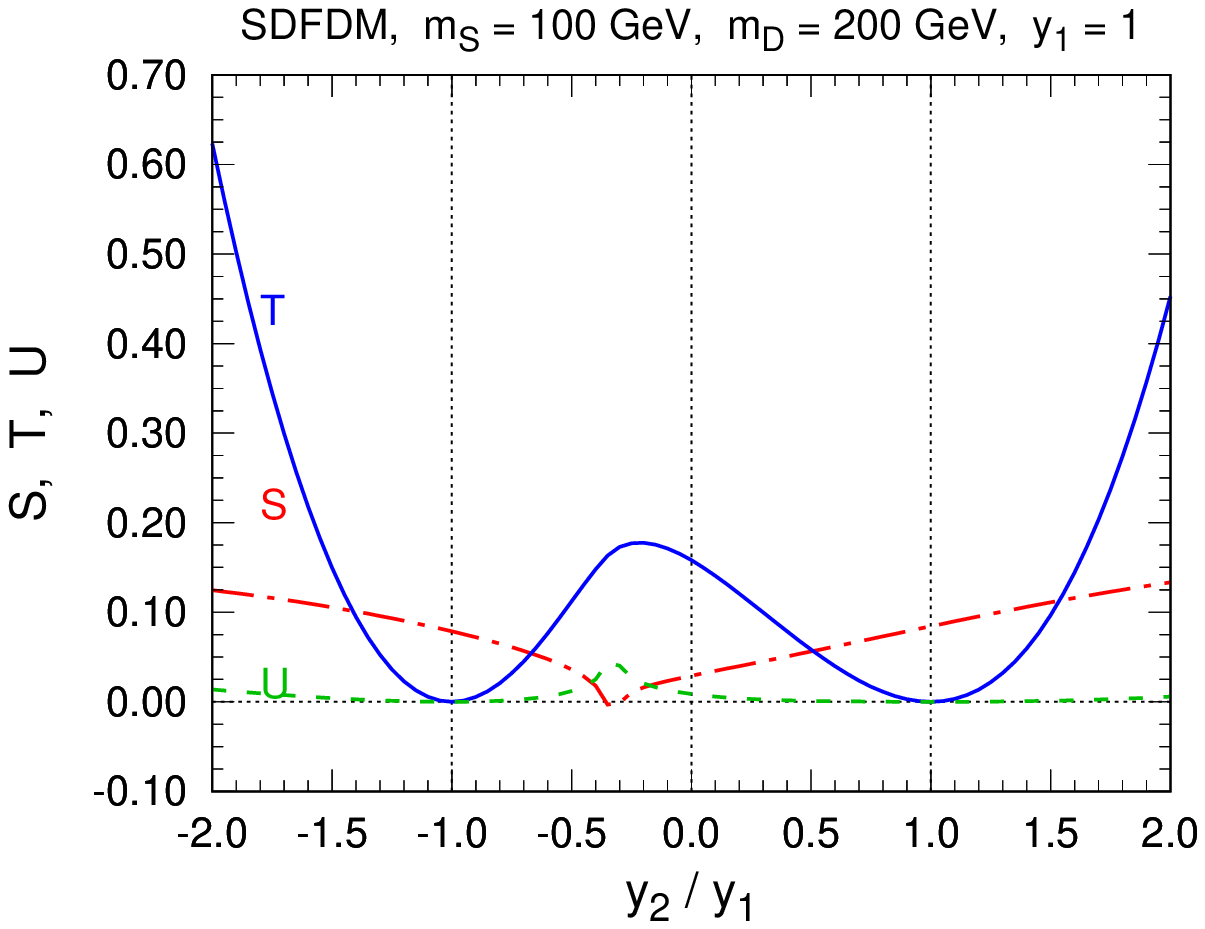}}
\subfigure[~$m_S>m_D$ case\label{SDSTU:b}]
{\includegraphics[width=0.49\textwidth]{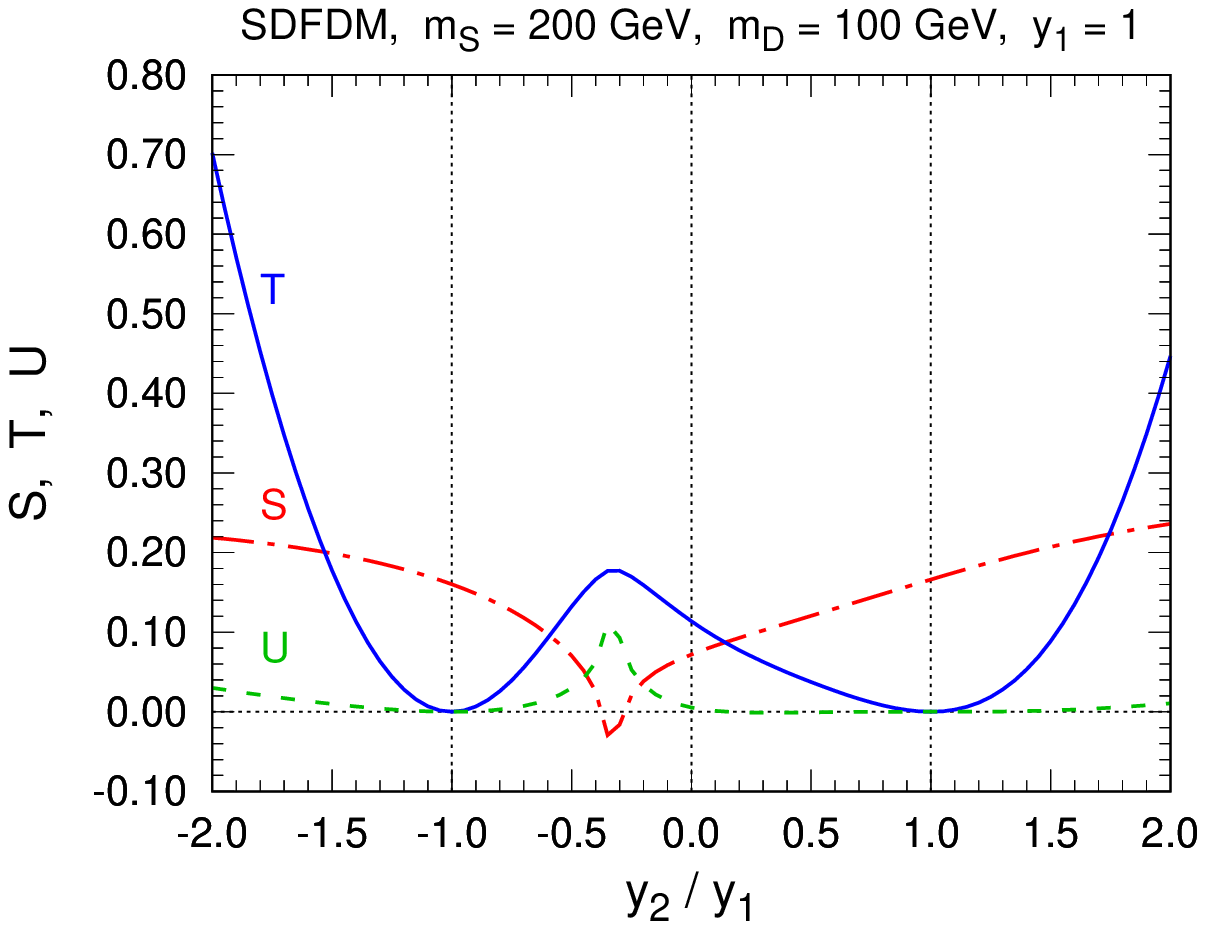}}
\caption{$S$, $T$, and $U$ as functions of $y_2/y_1$ in the SDFDM model  with $y_1=1$. In the left (right) panel, $m_S=100~(200)~\si{GeV}$ and  $m_D=200~(100)~\si{GeV}$.}
\label{SDSTU}
\end{figure}

Fig.~\ref{SDSTU} shows the $S$, $T$, and $U$ parameters as functions of the ratio $y_2/y_1$ in the SDFDM model with $y_1=1$. Two sets of $m_S$ and $m_D$ are chosen to separately represent the $m_S<m_D$ and $m_S>m_D$ cases.
In the custodial symmetry limits $y_2/y_1\to\pm 1$, $T$ and $U$ vanish as expected.
When $y_2/y_1\to\pm2$, $S$ and $T$ become large and will be strongly constrained by EW precision data.
$U$ is typically much smaller than the other two parameters, except for some special regions.

\begin{figure}[!t]
\centering
\subfigure[~$y_1=y_2=1$ (custodial symmetry)\label{SDS2d:a}]
{\includegraphics[width=0.49\textwidth]{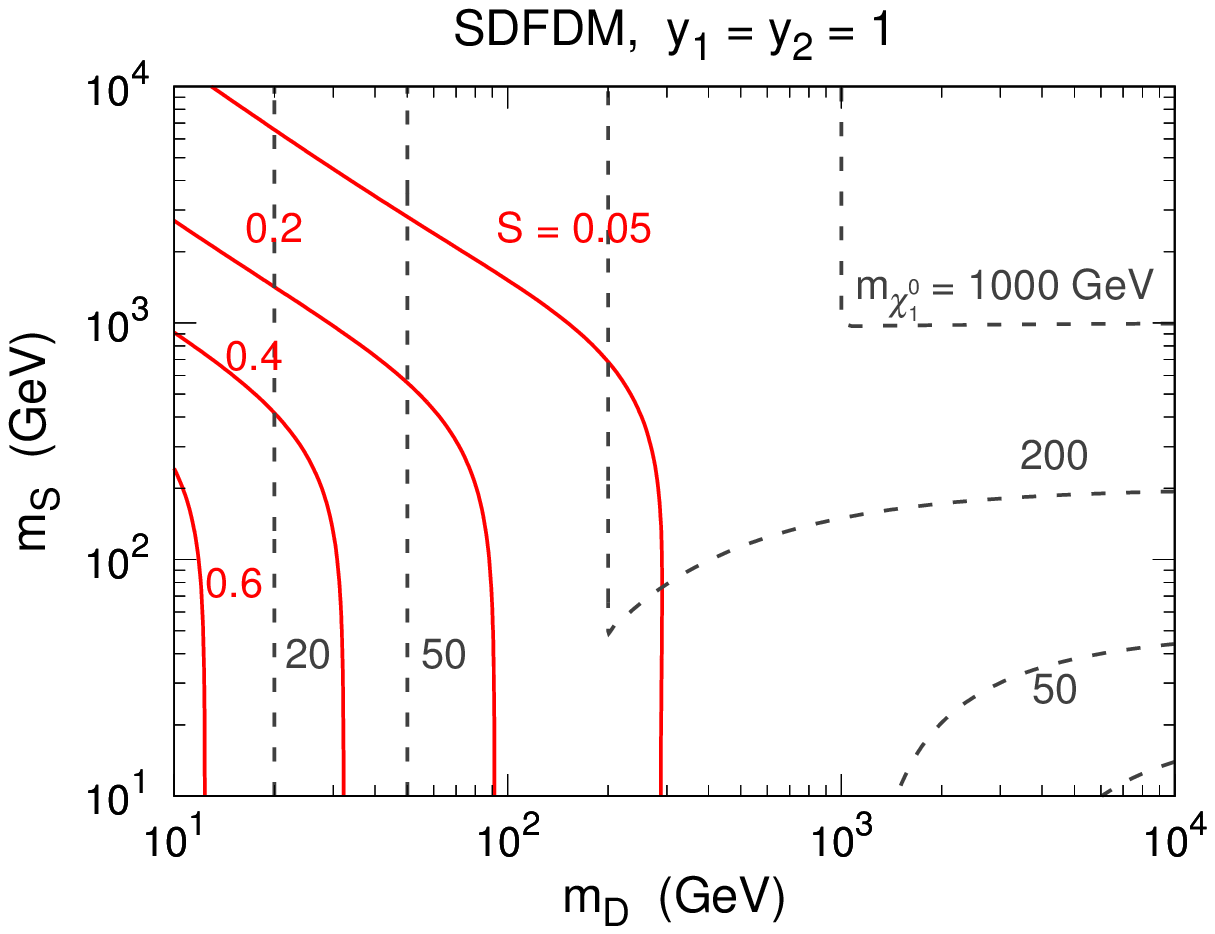}}
\subfigure[~$y_1=1$, $y_2=1.5$\label{SDS2d:b}]
{\includegraphics[width=0.49\textwidth]{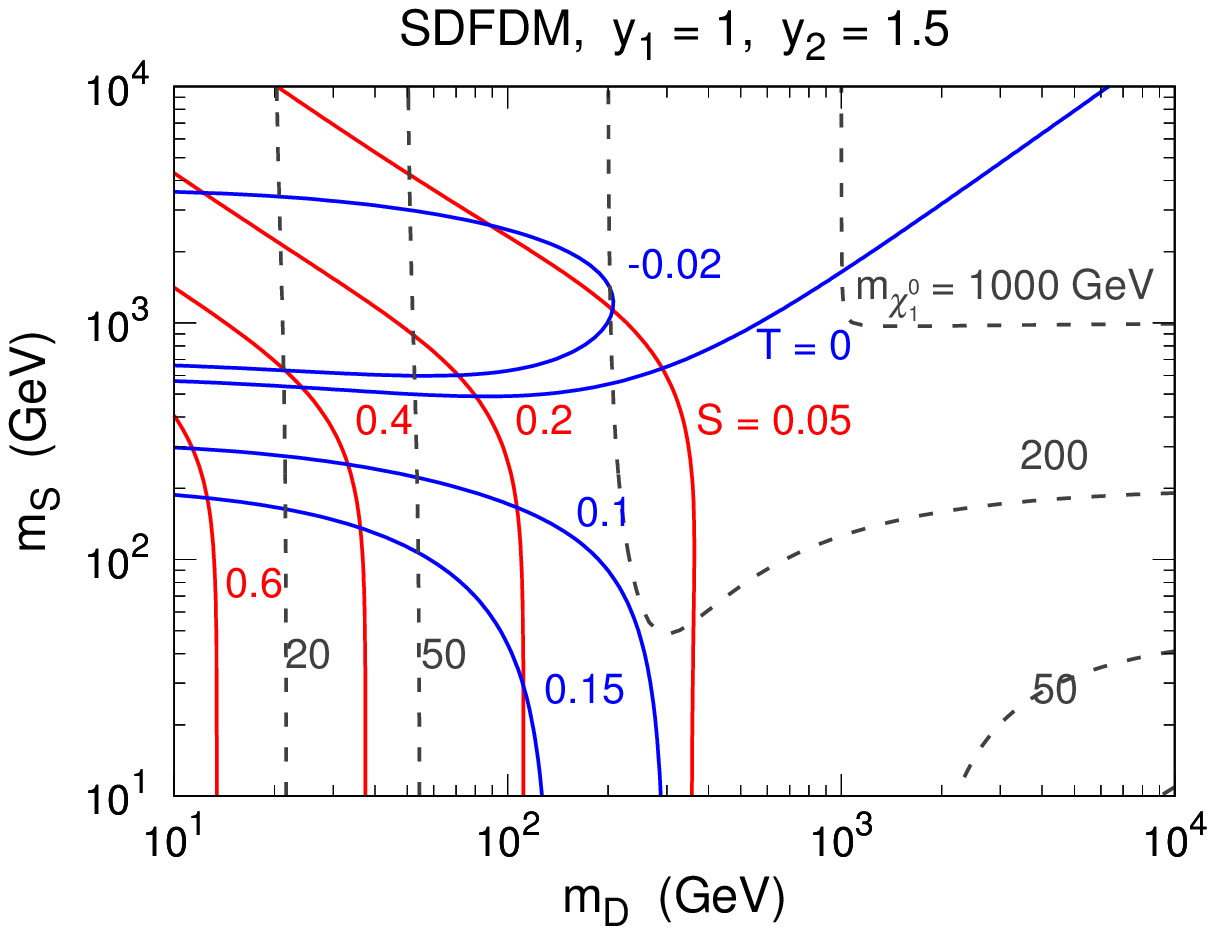}}
\caption{Contours of $S$ (red solid lines), $T$ (blue solid lines), and $m_{\chi_1^0}$ (gray dashed lines) in the $m_D-m_S$ plane for the SDFDM model with fixed $y_1$ and $y_2$.}
\label{SDS2d}
\end{figure}

In Fig.~\ref{SDS2d}, we show the contours of $S$, $T$, and $m_{\chi_1^0}$ in the $m_D-m_S$ plane.
Fig.~\ref{SDS2d:a} corresponds to the custodial symmetry, where $T$ and $U$ are always vanish, and only the behavior of $S$ are demonstrated.
In the region with small $m_S$ and $m_D$, $S$ is an $\mathcal{O}(0.1)$ number, decreasing as $m_D$ increases. This behavior can be understood as follows.
For $m_D<m_S\ll yv$ ($y=y_1=y_2$), the mass spectrum becomes $m_{\chi_1^0}=m_{\chi^\pm}= m_D$, $m_{\chi_{2,3}^0}\approx yv$. Using the expressions \eqref{eq:equalmass:J1}, \eqref{eq:equalmass:B0}, and \eqref{eq:appr:J1}, we have
\begin{equation}\label{eq:appr:S}
S\approx\frac{1}{\pi}\left[\frac{1}{2}J'_1(0,m_D^2,y^2v^2)-\frac{1}{2}J'_1(0,m_D^2,m_D^2)+m_D^2B'_0(0,m_D^2,m_D^2)\right]
\approx\frac{1}{\pi}\left(\frac{2}{3}\ln\frac{yv}{m_D}-\frac{1}{9}\right),
\end{equation}
where the leading term $\propto\ln(yv/m_D)$ and becomes smaller as $m_D$ increases.

Fig.~\ref{SDS2d:b} demonstrates the effect of custodial symmetry violation with $y_1=1$ and $y_2=1.5$.
$T$ is negative in the region where $m_S>m_D$ with $m_S\gtrsim 600$~GeV.
In the region with small $m_S$ and $m_D$, $T$ is positive and grow quickly as $m_D$ and $m_S$ decrease.

\begin{figure}[!t]
\centering
\subfigure[~$y_1=y_2=1$ (custodial symmetry)\label{SD2d:a}]
{\includegraphics[width=0.49\textwidth]{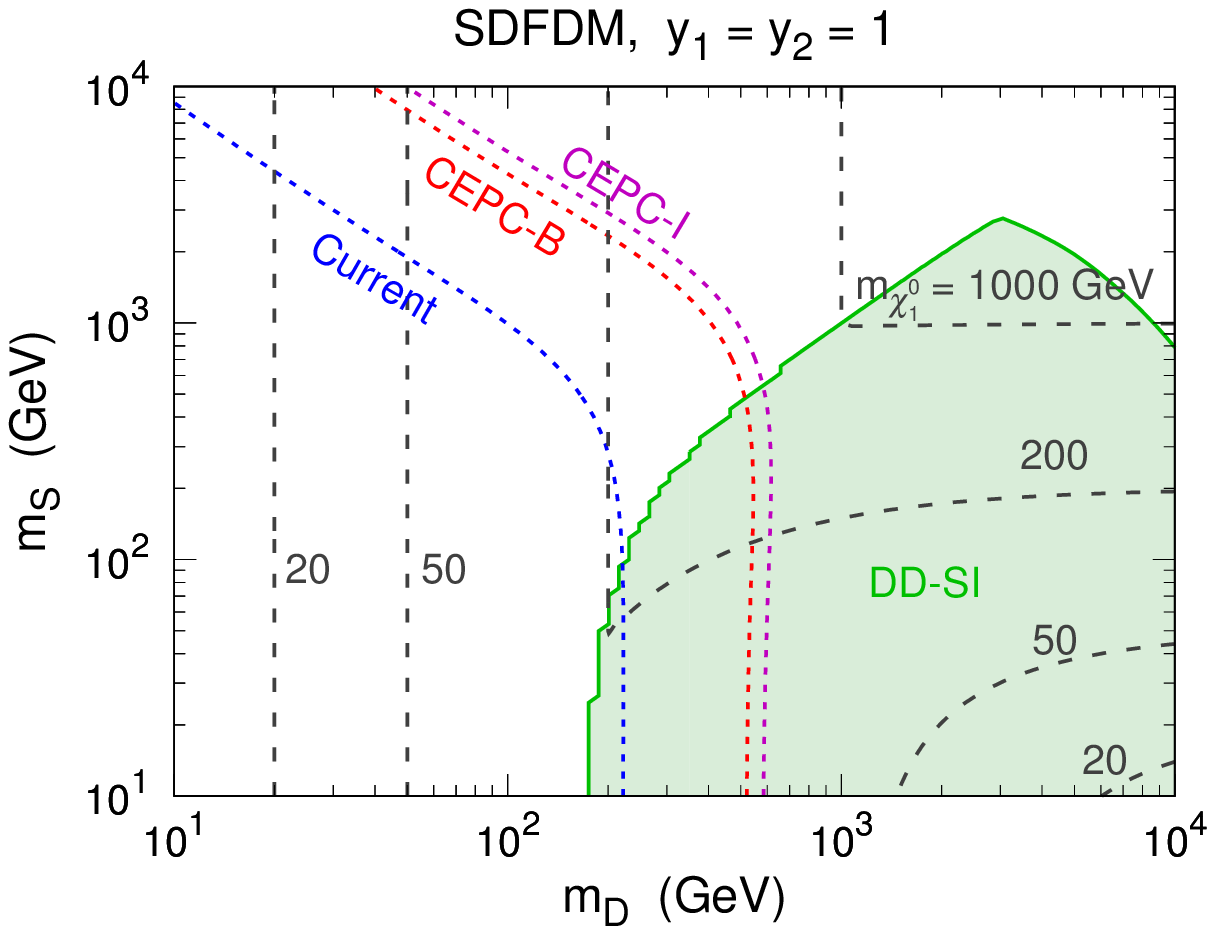}}
\subfigure[~$y_1=1$, $y_2=1.5$\label{SD2d:b}]
{\includegraphics[width=0.49\textwidth]{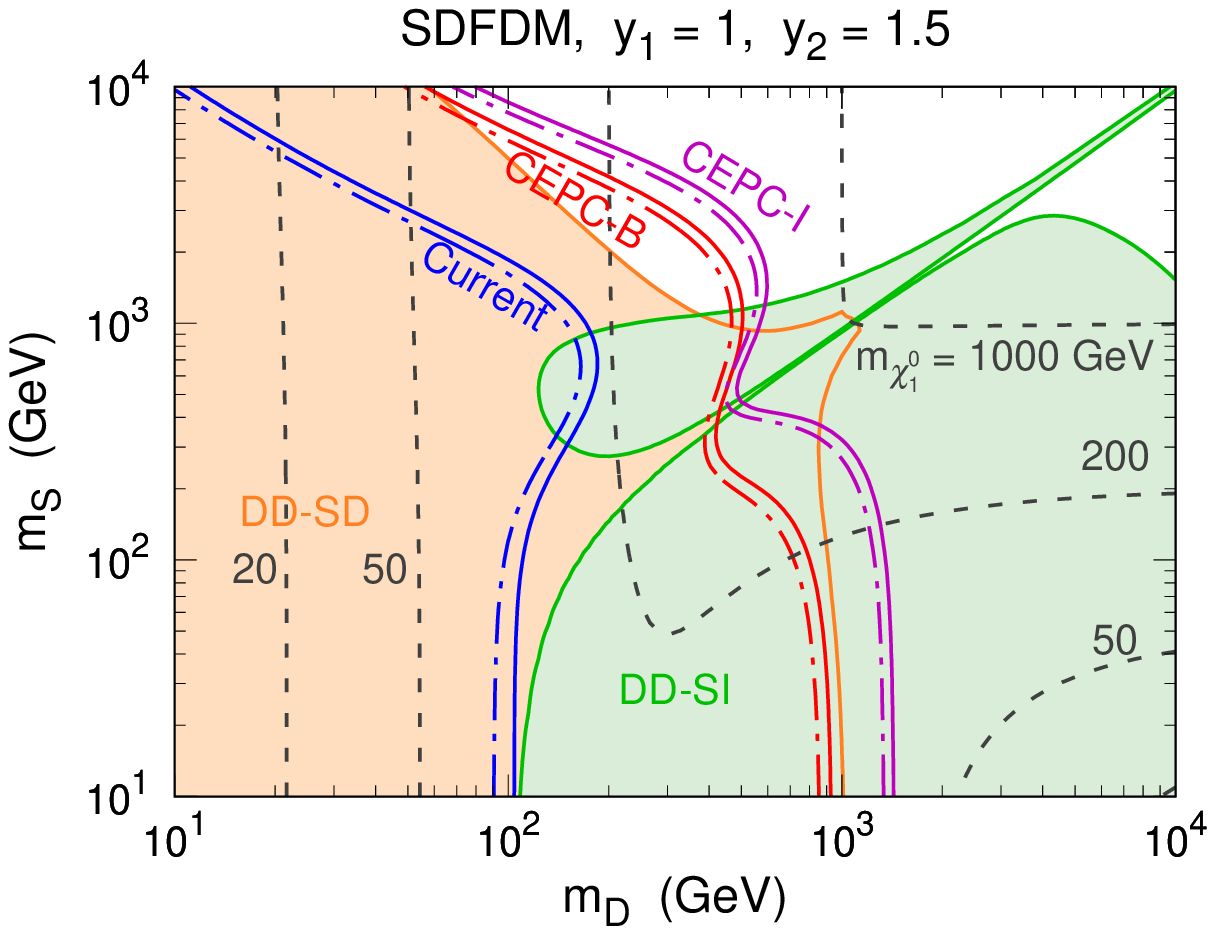}}
\caption{Expected 95\% CL constraints from CEPC precisions of EW oblique parameters in the $m_D-m_S$ plane for the SDFDM model with fixed $y_1$ and $y_2$. The constraints from current, CEPC-B, and CEPC-I precisions are denoted by blue, red, and purple colors, while dotted, solid, and dot-dashed lines correspond to the global fits with $T=U=0$, $U=0$, and no constraint on oblique parameters, respectively. The filled orange region is excluded by spin-dependent direct detection (DD-SD), while the filled green region is excluded by the spin-independent direct detection (DD-SI). Gray dashed lines indicate contours of the DM candidate mass $m_{\chi_1^0}$.}
\label{SD2d}
\end{figure}

In Fig.~\ref{SD2d}, we present the expected 95\% CL constraints in the $m_D-m_S$ plane from EW oblique parameters after the running of CEPC, as well as that from the current precision.
For the custodial symmetry limit $y_1=y_2$, we use the fit results obtained by assuming $T=U=0$ and denote the constraints by dotted lines.
The solid lines corresponds to the constraints from the global fits under the assumption of $U=0$, which should be a good approximation for the SDFDM model. The constraints from the global fits with free $S$, $T$, and $U$ are indicated by the dot-dashed lines and always weaker than the former constraints.

For comparison, we also show the constraints from direct detection experiments.
The green region denoted as ``DD-SI'' is excluded by the 90\% CL upper limits on the DM-nucleon SI scattering cross section from PandaX-II~\cite{Tan:2016zwf} and LUX~\cite{Akerib:2016vxi}.
Moreover, the orange region denoted as ``DD-SD'' is excluded at 90\% CL by the upper limit on the DM-neutron SD cross section from LUX~\cite{Akerib:2016lao} and the upper limits on the DM-proton SD cross section from PICO~\cite{Amole:2017dex}.

In Fig.~\ref{SD2d:a}, the Yukawa couplings are chosen to be $y_1=y_2=1$, respecting the custodial symmetry.
In this case, SD direct detection cannot put any bound since $g_{Z\chi_1^0\chi_1^0}=0$.
A large region for $m_D>m_S$ is excluded by SI direct detection.
On the other hand, the $m_D<m_S$ half plane evades this constraint because the $h\chi_1^0\chi_1^0$ coupling vanishes.
Nevertheless, current EW precision data can test this region up to $m_{\chi_1^0}\sim 200~\si{GeV}$, while the running of CEPC is expected to explore up to $m_{\chi_1^0}\sim 600~\si{GeV}$.

In Fig.~\ref{SD2d:b}, we fix $y_1=1$ and $y_2=1.5$, which do not respect the custodial symmetry.
We can find that SI and SD direct detection results collectively exclude a quite large region.
Even so, CEPC can still explore further in the parameter space, up to $m_{\chi_1^0}\sim 600~\si{GeV}$.

\section{Doublet-Triplet Fermionic Dark Matter}
\label{sec:DTFDM}

\subsection{Fields and Interactions}

In the DTFDM model, we consider a dark sector with two doublet and one triplet Weyl spinors:~\cite{Dedes:2014hga}
\begin{equation}
D_1=\begin{pmatrix}D_1^0\\D_1^-\end{pmatrix}\in(\mathbf{2},-1/2),\quad D_2=\begin{pmatrix}D_2^+\\D_2^0\end{pmatrix}\in(\mathbf{2},+1/2),\quad T=\begin{pmatrix}T^+\\T^0\\T^-\end{pmatrix}\in(\mathbf{3},0).
\end{equation}
The related Lagrangians are
\begin{eqnarray}
\mathcal{L}_\mathrm{D}&=&iD_{1}^\dag\bar{\sigma}^\mu D_\mu D_1+iD_{2}^\dag\bar{\sigma}^\mu D_\mu D_2+(m_D\epsilon_{ij}D_1^iD_2^j+\mathrm{h.c.}),\\
\mathcal{L}_\mathrm{T}&=&iT^\dag\bar{\sigma}^\mu D_\mu T-\frac{1}{2}(m_TT^aT^a+\mathrm{h.c.}),\\
\mathcal{L}_\mathrm{HDT}&=&y_1H_iT^a(\sigma^a)_j^iD_1^j-y_2H_i^\dag T^a(\sigma^a)_j^i D_2^j+\mathrm{h.c.}\,.
\end{eqnarray}
Triplet interactions with EW gauge bosons can be expressed as
\begin{eqnarray}\label{LT}
\mathcal{L}_\mathrm{T}&\supset&g[W^+_\mu(T^+)^\dag\bar{\sigma}^\mu T^0
+W^-_\mu(T^0)^\dag\bar{\sigma}^\mu T^+
-W^+_\mu(T^0)^\dag\bar{\sigma}^\mu T^-
-W^-_\mu(T^-)^\dag\bar{\sigma}^\mu T^0]\nonumber\\
&&+(eA_\mu+gc_\mathrm{W}Z_\mu)(T^+)^\dag\bar{\sigma}^\mu T^+
-(eA_\mu+gc_\mathrm{W}Z_\mu)(T^-)^\dag\bar{\sigma}^\mu T^-,
\end{eqnarray}
while gauge interactions of the doublets have been given by \eqref{LD}.

After the Higgs field develops a VEV, we have the mass terms
\begin{eqnarray}
\mathcal{L}_\mathrm{mass}&=&-\frac{1}{2}(T^0\quad D_1^0\quad D_2^0)\mathcal{M}_\mathrm{N}\begin{pmatrix}T^0\\D_1^0\\D_2^0\end{pmatrix}-(T^-\quad D_1^-)\mathcal{M}_\mathrm{C}\begin{pmatrix}T^+\\D_2^+\end{pmatrix}+\mathrm{h.c.}\nonumber\\
&=&-\frac{1}{2}\sum_{i=1}^3m_{\chi_i^0}\chi_i^0\chi_i^0-\sum_{i=1}^2m_{\chi^\pm_i}\chi^-_i\chi^+_i+\mathrm{h.c.}\,,
\end{eqnarray}
where the mass and mixing matrices are defined as
\begin{eqnarray}
&&\mathcal{M}_\mathrm{N}=\begin{pmatrix}m_T&\dfrac{1}{\sqrt{2}}y_1v&-\dfrac{1}{\sqrt{2}}y_2v\\[1em] \dfrac{1}{\sqrt{2}}y_1v&0&m_D\\[1em] -\dfrac{1}{\sqrt{2}}y_2v&m_D&0\end{pmatrix},\quad
\mathcal{M}_\mathrm{C}=\begin{pmatrix}m_T&-y_2v\\ -y_1v&-m_D\end{pmatrix},\\
&&\mathcal{N}^\mathrm{T}\mathcal{M}_\mathrm{N}\mathcal{N}=\mathrm{diag}(m_{\chi_1^0},m_{\chi_2^0},m_{\chi_3^0}),\quad
\mathcal{C}_\mathrm{R}^\mathrm{T}\mathcal{M}_\mathrm{C}\mathcal{C}_\mathrm{L}=\mathrm{diag}(m_{\chi_1^\pm},m_{\chi_2^\pm}),\\
&&\begin{pmatrix}T^0\\D_1^0\\D_2^0\end{pmatrix}=\mathcal{N}\begin{pmatrix}\chi_1^0\\ \chi_2^0\\ \chi_3^0\end{pmatrix},\quad
\begin{pmatrix}T^+\\D_2^+\end{pmatrix}=\mathcal{C}_\mathrm{L}\begin{pmatrix}\chi_1^+\\ \chi_2^+\end{pmatrix},\quad
\begin{pmatrix}T^-\\D_1^-\end{pmatrix}=\mathcal{C}_\mathrm{R}\begin{pmatrix}\chi_1^-\\ \chi_2^-\end{pmatrix}.
\end{eqnarray}

The dark sector involves three Majorana fermions $\chi_{1,2,3}^0$ and two singly charged fermion $\chi_{1,2}^\pm$.
The couplings of the DM candidate $\chi_1^0$ to the Higgs and $Z$ bosons are
\begin{equation}
g_{h\chi_1^0\chi_1^0}=-\sqrt{2}(y_1\mathcal{N}_{21}-y_2\mathcal{N}_{31})\mathcal{N}_{11},\quad
g_{Z\chi_1^0\chi_1^0}=-\frac{g}{2c_\mathrm{W}}(|\mathcal{N}_{21}|^2-|\mathcal{N}_{31}|^2).
\end{equation}
These couplings are related to direct detection.

\subsection{Vacuum Polarizations and Custodial Symmetry}

For evaluating the oblique parameters, we calculate the dark sector contributions to the vacuum polarizations:
\begin{eqnarray}
\Pi_{AA}(p^2)&=&\frac{2e^2}{16\pi^2}\sum_{i=1}^2J_2(p^2,m_{\chi^\pm_i}^2),\\
\Pi_{ZA}(p^2)&=&\frac{1}{16\pi^2}\sum_{i=1}^2e(a_{Z\chi_i^+\chi_i^-}+b_{Z\chi_i^+\chi_i^-})J_2(p^2,m_{\chi^\pm_i}^2),\\
\Pi_{ZZ}(p^2)&=&\frac{1}{16\pi^2}\bigg\{\sum_{i,j=1}^{3}\big[|g_{Z\chi_i^0\chi_j^0}|^2J_1(p^2,m_{\chi_i^0}^2,m_{\chi_j^0}^2)\nonumber\\
&&\qquad\qquad\qquad+m_{\chi_i^0}m_{\chi_j^0}(g_{Z\chi_i^0\chi_j^0}^2+\mathrm{h.c.})B_0(p^2,m_{\chi_i^0}^2,m_{\chi_j^0}^2)\big]\nonumber\\
&&\quad+\sum_{i,j=1}^2\big[(|a_{Z\chi_i^+\chi_j^-}|^2+|b_{Z\chi_i^+\chi_j^-}|^2)J_1(p^2,m_{\chi_i^\pm}^2,m_{\chi_j^\pm}^2)\nonumber\\
&&\qquad\qquad\quad-2m_{\chi_i^\pm}m_{\chi_j^\pm}(a_{Z\chi_j^+\chi_i^-}b_{Z\chi_i^+\chi_j^-}+\mathrm{h.c.})B_0(p^2,m_{\chi_i^\pm}^2,m_{\chi_j^\pm}^2)\big]\bigg\},\\
\Pi_{WW}(p^2)&=&\frac{1}{16\pi^2}\sum_{i=1}^3\sum_{j=1}^2\big[(|a_{W\chi_j^+\chi_i^0}|^2+|b_{W\chi_j^+\chi_i^0}|^2)J_1(p^2,m_{\chi_i^0}^2,m_{\chi_j^\pm}^2)\nonumber\\
&&\qquad\qquad\qquad-2m_{\chi_i^0}m_{\chi_j^\pm}(a_{W\chi_j^+\chi_i^0}b_{W\chi_j^+\chi_i^0}^\ast+\mathrm{h.c.})B_0(p^2,m_{\chi_i^0}^2,m_{\chi_j^\pm}^2)\big],\qquad
\end{eqnarray}
where the definitions of couplings are
\begin{eqnarray}
a_{Z\chi_i^+\chi_j^-}&=&gc_\mathrm{W}(\mathcal{C}_\mathrm{L})_{1,i}^\ast(\mathcal{C}_\mathrm{L})_{1,j}+\frac{g(c_\mathrm{W}^2-s_\mathrm{W}^2)}{2c_\mathrm{W}}(\mathcal{C}_\mathrm{L})_{2,i}^\ast(\mathcal{C}_\mathrm{L})_{2,j},\\
b_{Z\chi_i^+\chi_j^-}&=&gc_\mathrm{W}(\mathcal{C}_\mathrm{R})_{1,i}(\mathcal{C}_\mathrm{R})_{1,j}^\ast+\frac{g(c_\mathrm{W}^2-s_\mathrm{W}^2)}{2c_\mathrm{W}}(\mathcal{C}_\mathrm{R})_{2,i}(\mathcal{C}_\mathrm{R})_{2,j}^\ast,\\
g_{Z\chi^0_i\chi^0_j}&=&-\frac{g}{2c_\mathrm{W}}(\mathcal{N}_{2,i}\mathcal{N}_{2,j}^\ast-\mathcal{N}_{3,i}\mathcal{N}_{3,j}^\ast),\\
a_{W\chi_i^+\chi_j^0}&=&g(\mathcal{C}_\mathrm{L})_{1,i}^\ast\mathcal{N}_{1,j}+\frac{g}{\sqrt{2}}(\mathcal{C}_\mathrm{L})_{2,i}^\ast\mathcal{N}_{3,j},\quad
b_{W\chi_i^+\chi_j^0}=g(\mathcal{C}_\mathrm{R})_{1,i}\mathcal{N}_{1,j}^\ast-\frac{g}{\sqrt{2}}(\mathcal{C}_\mathrm{R})_{2,i}\mathcal{N}_{2,j}^\ast.\qquad\quad
\end{eqnarray}

Analogous to the SDFDM model, the custodial symmetry exists if $y_1=\pm y_2$, leading to $T=U=0$ and $g_{Z\chi_1^0\chi_1^0}=0$. For instance, when $y_1=y_2=y$, we can define $\mathrm{SU}(2)_\mathrm{R}$ doublets
\begin{equation}
(\mathcal{D}^A)^i=\begin{pmatrix}D_1^i\\ D_2^i\end{pmatrix},\quad
(\mathcal{H}^A)^i=\begin{pmatrix}H_i^\dag\\ H_i\end{pmatrix},
\end{equation}
and obtain the $\mathrm{SU}(2)_\mathrm{L}\times\mathrm{SU}(2)_\mathrm{R}$ invariant Lagrangians
\begin{eqnarray}
\mathcal{L}_\mathrm{D}&=&i\mathcal{D}_A^\dag\bar{\sigma}^\mu D_\mu\mathcal{D}^A+\frac{1}{2}[m_D\epsilon_{AB}\epsilon_{ij}(\mathcal{D}^A)^i(\mathcal{D}^B)^j+\mathrm{h.c.}],\\
\mathcal{L}_\mathrm{HDT}&=&y\epsilon_{AB}(\mathcal{H}^A)_iT^a(\sigma^a)^i_j(\mathcal{D}^B)^j+\mathrm{h.c.}\,.
\end{eqnarray}
If $m_D<m_T$, we also have $g_{h\chi_1^0\chi_1^0}=0$.

\subsection{Expected Constraints}

\begin{figure}[!t]
\centering
\subfigure[~$m_D<m_T$ case\label{DTSTU:a}]
{\includegraphics[width=0.49\textwidth]{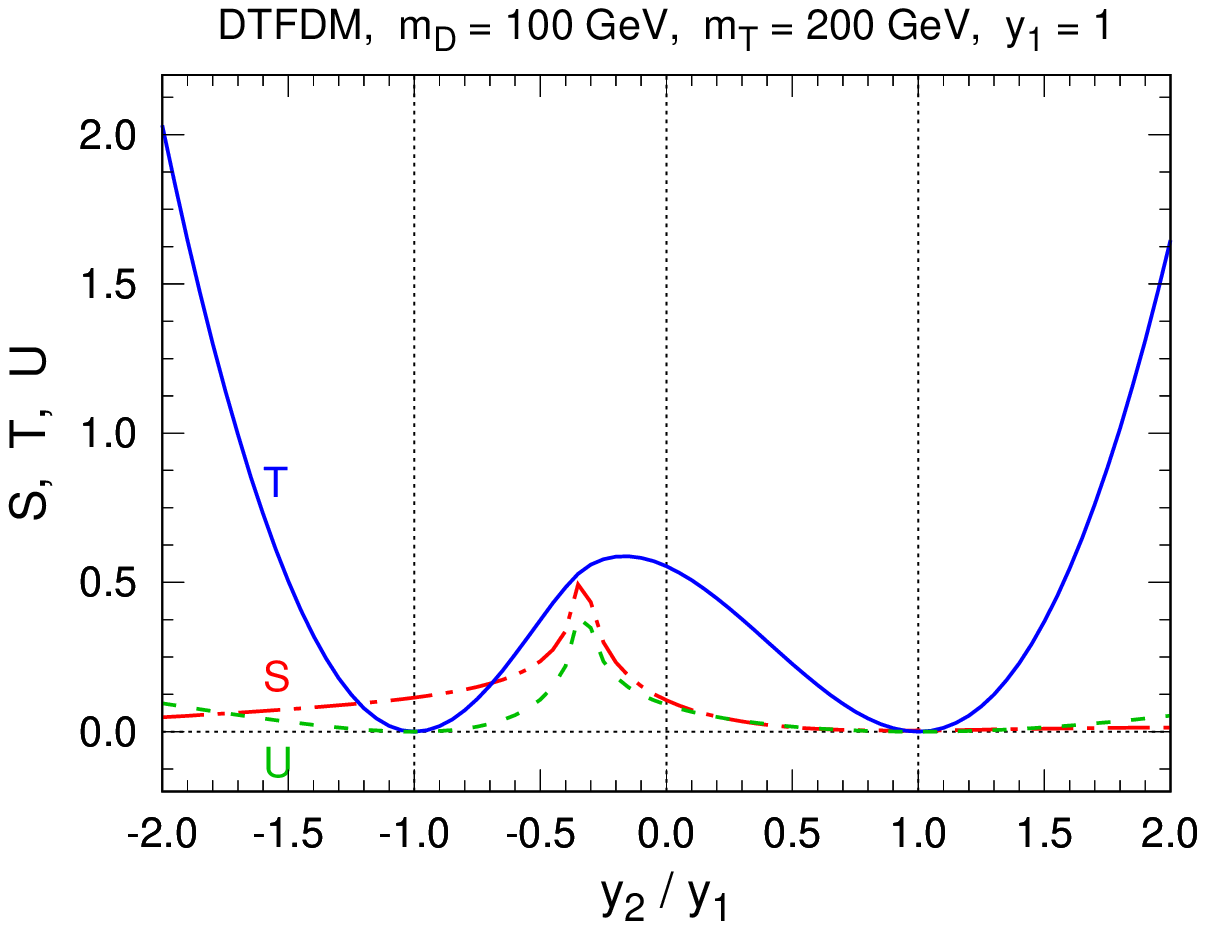}}
\subfigure[~$m_D>m_T$ case\label{DTSTU:b}]
{\includegraphics[width=0.49\textwidth]{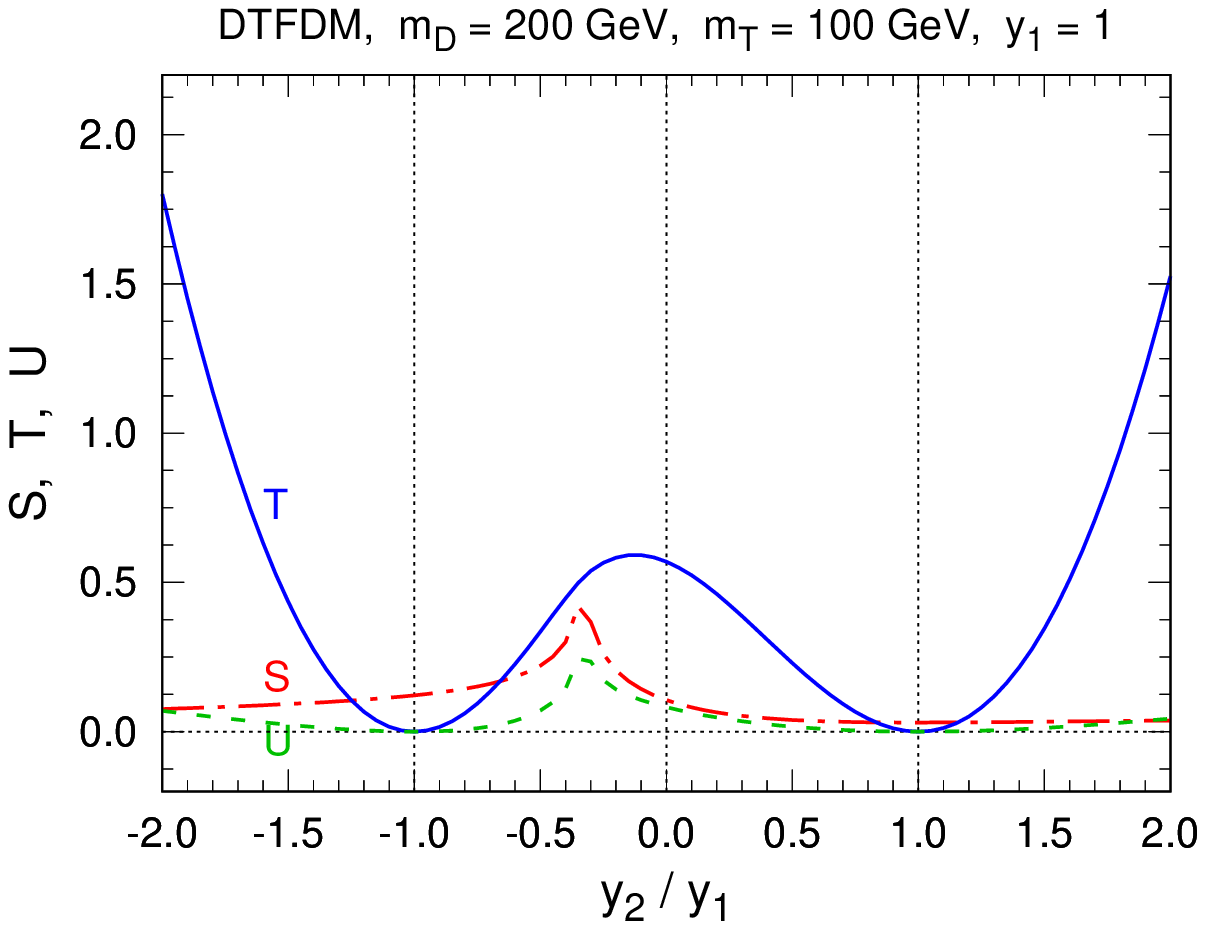}}
\caption{$S$, $T$, and $U$ as functions of $y_2/y_1$ in the DTFDM model  with $y_1=1$. In the left (right) panel, $m_D=100~(200)~\si{GeV}$ and  $m_T=200~(100)~\si{GeV}$.}
\label{DTSTU}
\end{figure}

In Fig.~\ref{DTSTU}, we show the EW oblique parameters as functions of $y_2/y_1$ in the DTFDM model.
$T$ and $U$ vanish at the points respecting the custodial symmetry, \textit{i.e.}, at $y_2/y_1=\pm 1$.
When this symmetry is violated, $T$ increases quickly.
$S$ tends to 0 for $0.5<y_2/y_1<2$ in both the $m_D<m_T$ and $m_T<m_D$ cases.

\begin{figure}[!t]
\centering
\subfigure[~$y_1=y_2=1$ (custodial symmetry)\label{DTS2d:a}]
{\includegraphics[width=0.49\textwidth]{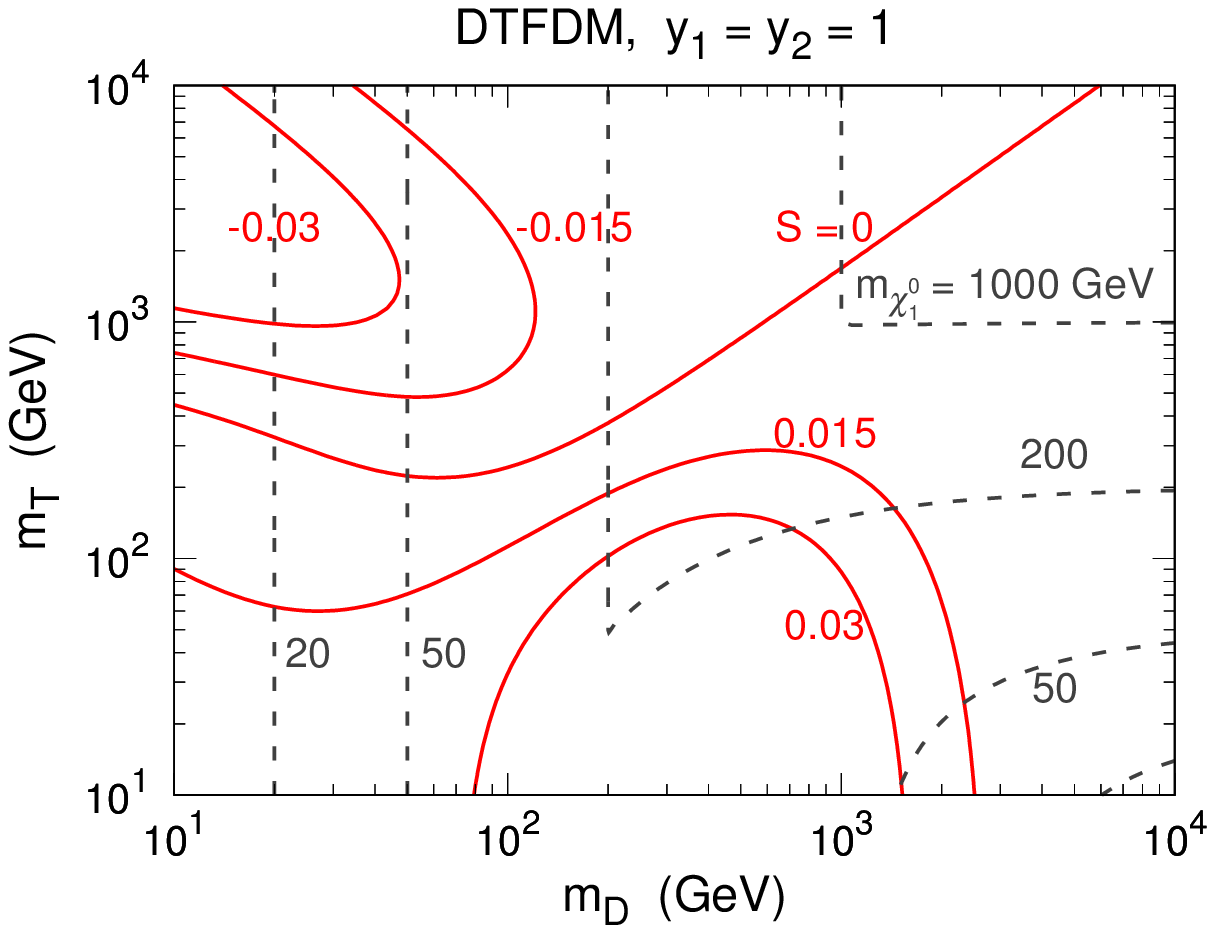}}
\subfigure[~$y_1=1$, $y_2=1.5$\label{DTS2d:b}]
{\includegraphics[width=0.49\textwidth]{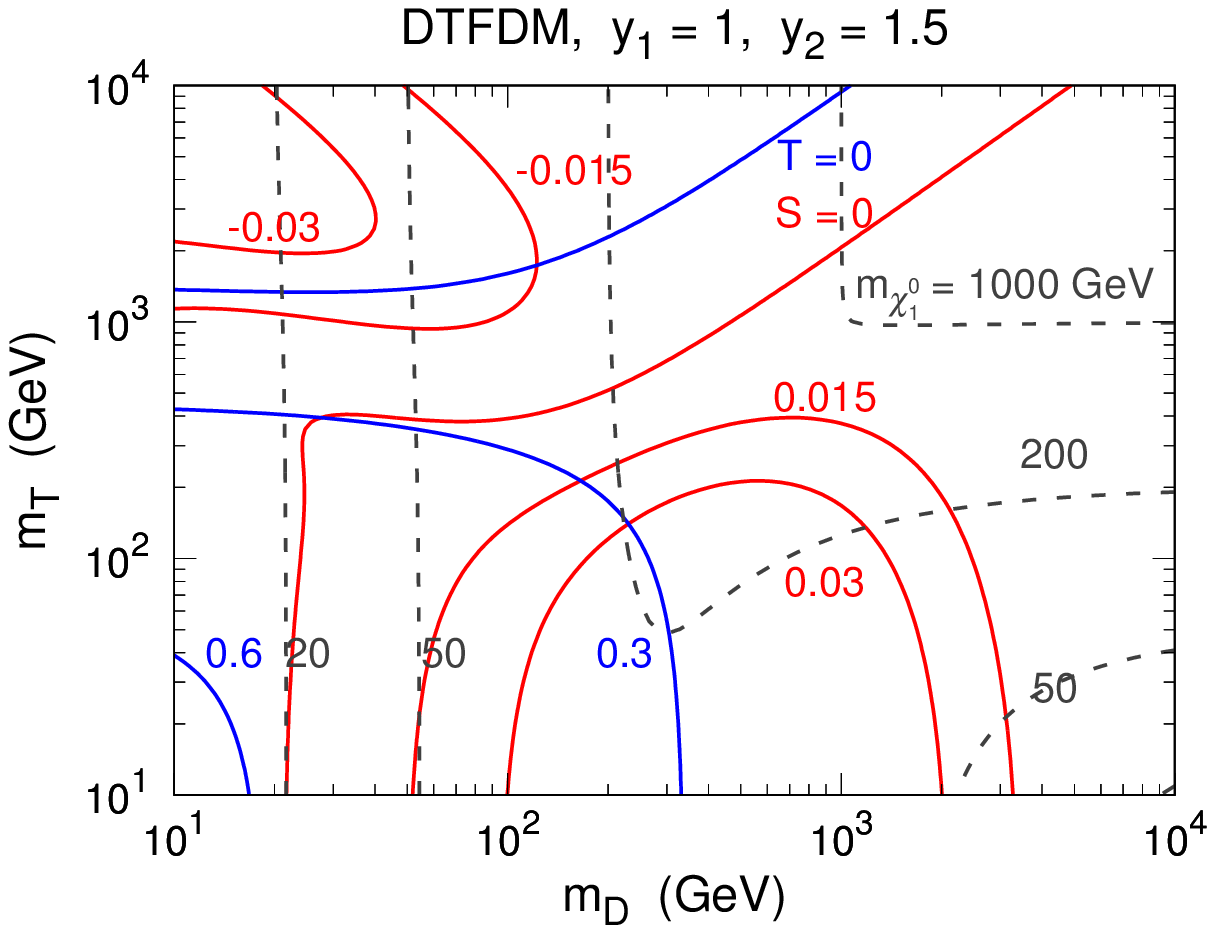}}
\caption{Contours of $S$ (red solid lines), $T$ (blue solid lines), and $m_{\chi_1^0}$ (gray dashed lines) in the $m_D-m_T$ plane for the DTFDM model with fixed $y_1$ and $y_2$.}
\label{DTS2d}
\end{figure}

In Fig.~\ref{DTS2d}, we further present the contours of $S$ and $T$ in the $m_D-m_T$ plane.
In contrast to the SDFDM model, $|S|\lesssim\mathcal{O}(0.01)$ in these plots, and there are contours corresponding to $S=0$, separating the regions with different signs of $S$.
An $S$ of $\mathcal{O}(0.01)$ is beyond the reach of current measurements, calling for future CEPC data.
For the custodial symmetry limit $y=y_1=y_2=1$, as shown in Fig.~\ref{DTS2d:a}, the masses of dark sector fermions in the region with $m_D<m_T\ll yv$ are $m_{\chi_1^0}= m_D$ and $m_{\chi_{2,3}^0}\approx m_{\chi_{1,2}^\pm}\approx yv$. Thus, we have
\begin{equation}
S\approx\frac{1}{\pi}\left[\frac{1}{2}J'_1(0,m_D^2,y^2v^2)-\frac{1}{2}J'_1(0,y^2v^2,y^2v^2)+2y^2v^2B'_0(0,y^2v^2,y^2v^2)\right]\approx\frac{1}{18\pi},
\end{equation}
which explains the $\mathcal{O}(0.01)$ magnitude of $S$.

The key to obtain the approximation \eqref{eq:appr:S} in the SDSDM model is that there is an unmixed charged particle which has a mass $m_D$.
This brings us a $J_1(0,m_D^2,m_D^2)$ term which contains a $\ln(m_D^2/Q^2)$ contribution to $\Pi_{ZZ}$. After taking into account the first term $J'_1(0,m_D^2,y^2v^2)$ that involves a $\ln(y^2v^2/Q^2)$ term, we have a significant $\ln(y^2v^2/m_D^2)$ contribution to $S$ in the end.
On the order hand, the two charged fermions in the DTFDM model mix up and both their masses tend to $yv$ when $m_D\ll yv$.
Consequently, their contribution to $\Pi_{ZZ}$ leads to a $J_1(0,y^2v^2,y^2v^2)$ term, which involve a $\ln(y^2v^2/Q^2)$ term that is canceled by the first term $J'_1(0,m_D^2,y^2v^2)$.
Therefore, there is no significant logarithmic contribution any more, leading to a much smaller $S$

Fig.~\ref{DTS2d:b} corresponds the case violating the custodial symmetry, where the $T$ parameter turns on.
We fix $y_1=1$ and $y_2=1.5$ fixed and find that $T$ is negative in a region where $m_T>m_D$ and $m_T\gtrsim 1~\si{TeV}$.
For small $m_D$ and $m_T$, $T$ is positive and grows fast as $m_D$ and $m_T$ decrease.
For large $m_D$ and $m_T$, $S$ has similar values to the custodial symmetric case. For very small mass parameters, however, it becomes negative.

\begin{figure}[!t]
\centering
\subfigure[~$y_1=y_2=1$ (custodial symmetry)\label{DT2d:a}]
{\includegraphics[width=0.49\textwidth]{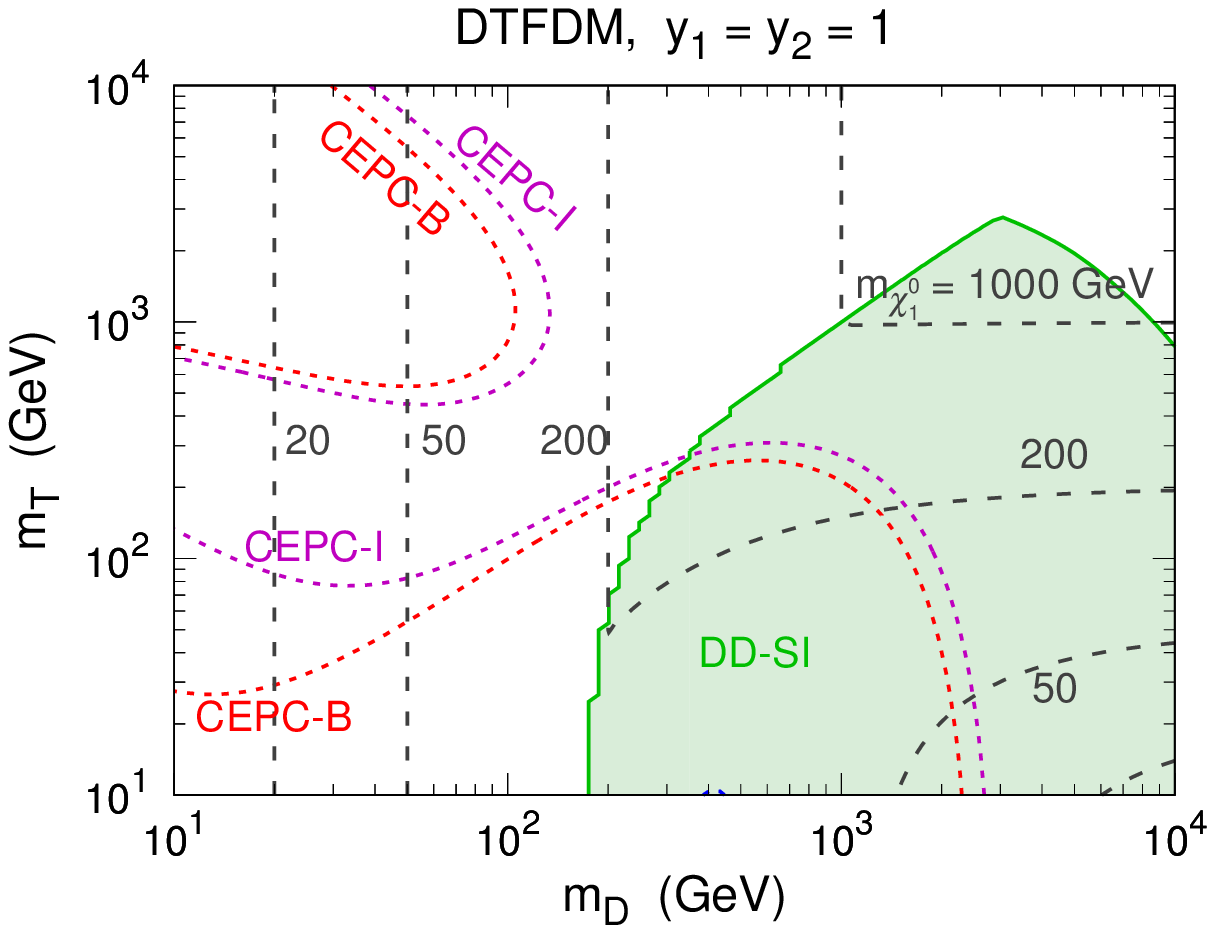}}
\subfigure[~$y_1=1$, $y_2=1.5$\label{DT2d:b}]
{\includegraphics[width=0.49\textwidth]{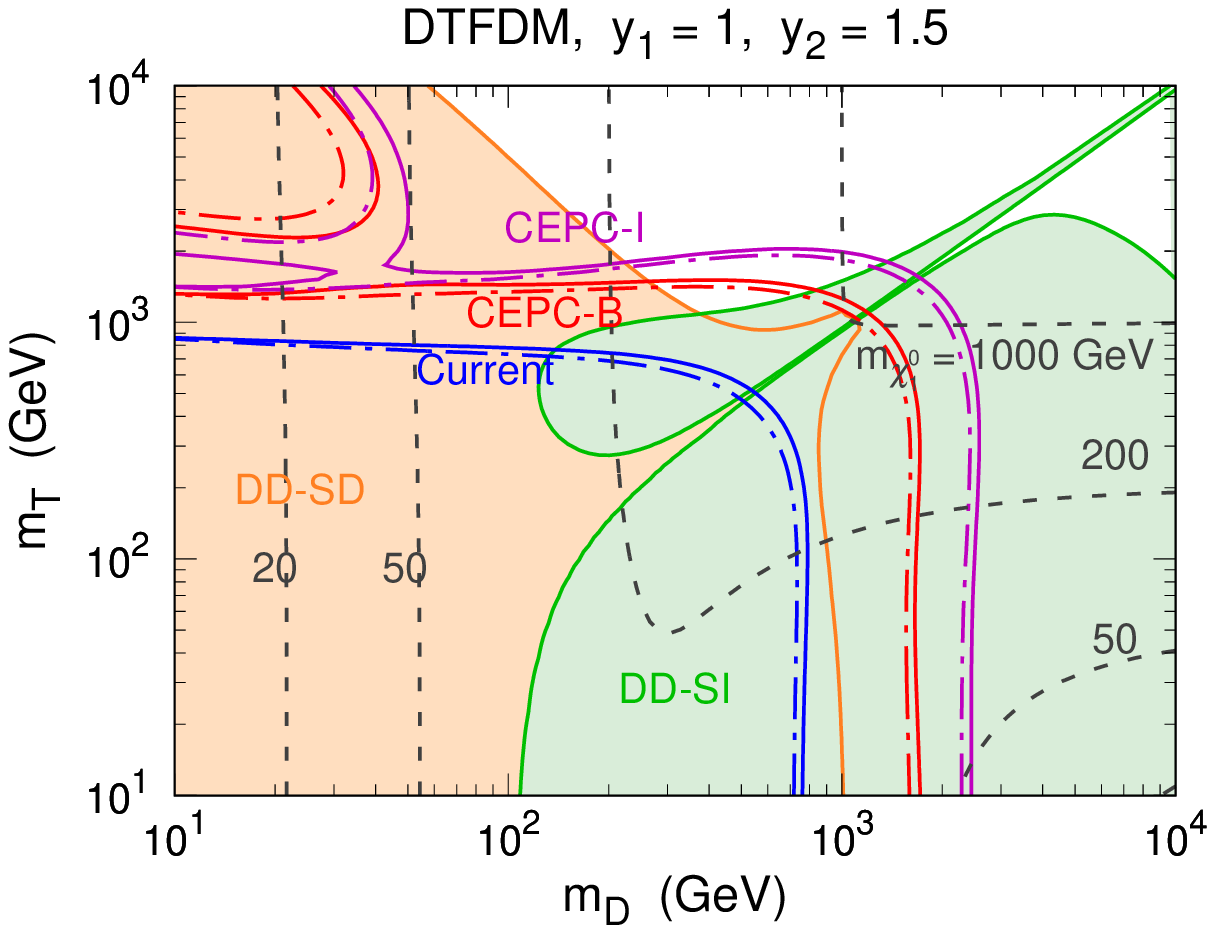}}
\caption{Expected 95\% CL constraints from CEPC precisions of EW oblique parameters in the $m_D-m_T$ plane for the DTFDM model with fixed $y_1$ and $y_2$. The meanings of labels, colors, and line types are the same as in Fig.~\ref{SD2d}.}
\label{DT2d}
\end{figure}

Expected constraints from the CEPC determination of EW oblique parameters and current bounds from direct detection on the DTFDM model in the $m_D-m_T$ plane are shown in Fig.~\ref{DT2d}.
The DD-SI and DD-SD exclusion regions are very similar to those in the SDFDM model.
Nevertheless, for the reason discussed above, the expected constraints from EW data are quite different.
As shown in Fig.~\ref{DT2d:a} for the custodial symmetry limit $y_1=y_2=1$, current EW precision measurements are not sensitive at all. CEPC data are sensitive to two separate regions, but a large portion of parameter space with moderate mass parameters cannot be explored.
For the case with $y_1=1$ and $y_2=1.5$ in Fig.~\ref{DT2d:b}, CEPC measurements can probe up to $m_{\chi_1^0}\sim 2~\si{TeV}$, but only a small portion of the CEPC-sensitive region is not excluded by current direct experiments.

\section{Triplet-Quadruplet Fermionic Dark Matter}
\label{sec:TQFDM}

\subsection{Fields and Interactions}

In the TQFDM model, one triplet and two quadruplet Weyl spinors are introduced:~\cite{Tait:2016qbg}
\begin{equation}
T=\begin{pmatrix}T^+\\T^0\\T^-\end{pmatrix}\in(\mathbf{3},0),\quad Q_1=\begin{pmatrix}Q_1^+\\Q_1^0\\Q_1^-\\Q_1^{--}\end{pmatrix}\in(\mathbf{4},-1/2),\quad Q_2=\begin{pmatrix}Q_2^{++}\\Q_2^+\\Q_2^0\\Q_2^-\end{pmatrix}\in(\mathbf{4},+1/2).
\end{equation}
Their properties are described by the Lagrangians
\begin{eqnarray}
\mathcal{L}_\mathrm{T}&=&iT^\dag\bar{\sigma}^\mu D_\mu T-\frac{1}{2}(m_TT^aT^a+\mathrm{h.c.}),\\
\mathcal{L}_\mathrm{Q}&=&iQ_{1}^\dag\bar{\sigma}^\mu D_\mu Q_1+iQ_{2}^\dag\bar{\sigma}^\mu D_\mu Q_2-(m_QQ_1Q_2+\mathrm{h.c.}),\\
\mathcal{L}_\mathrm{HTQ}&=&y_1Q_1TH-y_2Q_2TH^\dag+\mathrm{h.c.}\,,
\end{eqnarray}
where
\begin{eqnarray}
Q_1TH&=&\epsilon_{jl}(Q_1)_i^{jk}T^i_kH^l
\to\left(\frac{1}{\sqrt{3}}Q_1^-T^+-\frac{2}{\sqrt{6}}Q_1^0T^0-Q_1^+T^-\right)\frac{v+h}{\sqrt{2}},\\
Q_2TH^\dag&=&(Q_1)_i^{jk}T^i_kH^\dag_j
\to\left(-\frac{2}{\sqrt{6}}Q_2^0T^0-\frac{1}{\sqrt{3}}Q_2^+T^-+Q_2^-T^+\right)\frac{v+h}{\sqrt{2}}.
\end{eqnarray}
Gauge interactions of the quadruplets can be derived as
\begin{eqnarray}
\mathcal{L}_\mathrm{Q}&\supset&
\bigg\{\frac{\sqrt{6}}{2}gW^+_\mu[(Q_1^+)^\dag\bar{\sigma}^\mu Q_1^0+(Q_2^{++})^\dag\bar{\sigma}^\mu Q_2^+
+(Q_1^-)^\dag\bar{\sigma}^\mu Q_1^{--}+(Q_2^0)^\dag\bar{\sigma}^\mu Q_2^-]\nonumber\\
&&\quad+\sqrt{2}gW^+_\mu[(Q_1^0)^\dag\bar{\sigma}^\mu Q_1^-+(Q_2^+)^\dag\bar{\sigma}^\mu Q_2^0]+\mathrm{h.c.}\bigg\}\nonumber\\
&&+\left[eA_\mu+\frac{g(s_\mathrm{W}^2+3c_\mathrm{W}^2)}{2c_\mathrm{W}}Z_\mu\right](Q_1^+)^\dag\bar{\sigma}^\mu Q_1^+
+\left[eA_\mu+\frac{g(c_\mathrm{W}^2-s_\mathrm{W}^2)}{2c_\mathrm{W}}Z_\mu\right](Q_2^+)^\dag\bar{\sigma}^\mu Q_2^+
\nonumber\\
&&+\frac{g}{2c_\mathrm{W}}Z_\mu(Q_1^0)^\dag\bar{\sigma}^\mu Q_1^0
+\left[2eA_\mu+\frac{g(3c_\mathrm{W}^2-s_\mathrm{W}^2)}{2c_\mathrm{W}}Z_\mu\right](Q_2^{++})^\dag\bar{\sigma}^\mu Q_2^{++}\nonumber\\
&&-\left[eA_\mu+\frac{g(3c_\mathrm{W}^2+s_\mathrm{W}^2)}{2c_\mathrm{W}}Z_\mu\right](Q_2^-)^\dag\bar{\sigma}^\mu Q_2^-
-\left[eA_\mu+\frac{g(c_\mathrm{W}^2-s_\mathrm{W}^2)}{2c_\mathrm{W}}Z_\mu\right](Q_1^-)^\dag\bar{\sigma}^\mu Q_1^-\nonumber\\
&&-\frac{g}{2c_\mathrm{W}}Z_\mu(Q_2^0)^\dag\bar{\sigma}^\mu Q_2^0
-\left[2eA_\mu+\frac{g(3c_\mathrm{W}^2-s_\mathrm{W}^2)}{2c_\mathrm{W}}Z_\mu\right](Q_1^{--})^\dag\bar{\sigma}^\mu Q_1^{--}.
\end{eqnarray}
Gauge interactions of the triplet are the same as in \eqref{LT}.

Mass terms in the dark sector are
\begin{eqnarray}
\mathcal{L}_\mathrm{mass}&=&-\frac{1}{2}(T^0\quad Q_1^0\quad Q_2^0)\mathcal{M}_\mathrm{N}\begin{pmatrix}T^0\\Q_1^0\\Q_2^0\end{pmatrix}-(T^-\quad Q_1^-\quad Q_2^-)\mathcal{M}_\mathrm{C}\begin{pmatrix}T^+\\Q_1^+\\Q_2^+\end{pmatrix}\nonumber\\
&&-m_{Q}Q_1^{--}Q_2^{++}+\mathrm{h.c.}\nonumber\\
&=&-\frac{1}{2}\sum_{i=1}^3m_{\chi_i^0}\chi_i^0\chi_i^0-\sum_{i=1}^3m_{\chi^\pm_i}\chi^-_i\chi^+_i-m_{\chi^{\pm\pm}}\chi^{--}\chi^{++}+\mathrm{h.c.}\,,
\end{eqnarray}
where $m_{\chi^{\pm\pm}}=m_{Q}$, $\chi^{--}=Q_1^{--}$, $\chi^{++}=Q_2^{++}$, and the definitions of the mass and mixing matrices are
\begin{eqnarray}
&&\mathcal{M}_\mathrm{N}=\begin{pmatrix}m_T&\dfrac{1}{\sqrt{3}}y_1v&-\dfrac{1}{\sqrt{3}}y_2v\\[1em] \dfrac{1}{\sqrt{3}}y_1v&0&m_Q\\[1em] -\dfrac{1}{\sqrt{3}}y_2v&m_Q&0\end{pmatrix},\quad
\mathcal{M}_\mathrm{C}=\begin{pmatrix}m_T&\dfrac{1}{\sqrt{2}}y_1v&-\dfrac{1}{\sqrt{6}}y_2v\\[1em] -\dfrac{1}{\sqrt{6}}y_1v&0&-m_Q\\[1em]\dfrac{1}{\sqrt{2}}y_2v&-m_Q&0\end{pmatrix},\quad\\
&&\mathcal{N}^\mathrm{T}\mathcal{M}_\mathrm{N}\mathcal{N}=\mathrm{diag}(m_{\chi_1^0},m_{\chi_2^0},m_{\chi_3^0}),\quad
\mathcal{C}_\mathrm{R}^\mathrm{T}\mathcal{M}_\mathrm{C}\mathcal{C}_\mathrm{L}=\mathrm{diag}(m_{\chi_1^\pm},m_{\chi_2^\pm},m_{\chi_2^\pm}),\\
&&\begin{pmatrix}T^0\\Q_1^0\\Q_2^0\end{pmatrix}=\mathcal{N}\begin{pmatrix}\chi_1^0\\ \chi_2^0\\ \chi_3^0\end{pmatrix},\quad
\begin{pmatrix}T^+\\Q_1^+\\Q_2^+\end{pmatrix}=\mathcal{C}_\mathrm{L}\begin{pmatrix}\chi_1^+\\ \chi_2^+\\ \chi_3^+\end{pmatrix},\quad
\begin{pmatrix}T^-\\Q_1^-\\Q_2^-\end{pmatrix}=\mathcal{C}_\mathrm{R}\begin{pmatrix}\chi_1^-\\ \chi_2^-\\ \chi_3^-\end{pmatrix}.
\end{eqnarray}
There are three Majorana fermions $\chi_{1,2,3}^0$, three singly charged fermions $\chi_{1,2,3}^\pm$, and one doubly charged fermion $\chi^{\pm\pm}$.
The DM candidate $\chi_1^0$ has trilinear couplings to the Higgs and $Z$ bosons:
\begin{equation}
g_{h\chi_1^0\chi_1^0}=-\frac{2}{\sqrt{3}}(y_1\mathcal{N}_{21}-y_2\mathcal{N}_{31})\mathcal{N}_{11},\quad
g_{Z\chi_1^0\chi_1^0}=-\frac{g}{2c_\mathrm{W}}(|\mathcal{N}_{21}|^2-|\mathcal{N}_{31}|^2).
\end{equation}
Therefore, it may induce signals in direct detection experiments.

\subsection{Vacuum Polarizations and Custodial Symmetry}

The vacuum polarizations of EW gauge bosons contributed by dark sector fermions can be expressed as
\begin{eqnarray}
\Pi_{AA}(p^2)&=&\frac{2e^2}{16\pi^2}\sum_{i=1}^3J_2(p^2,m_{\chi^\pm_i}^2)+\frac{8e^2}{16\pi^2}J_2(p^2,m_{\chi^{\pm\pm}}^2),\\
\Pi_{ZA}(p^2)&=&\frac{1}{16\pi^2}\sum_{i=1}^3e(a_{Z\chi_i^+\chi_i^-}+b_{Z\chi_i^+\chi_i^-})J_2(p^2,m_{\chi^\pm_i}^2)
+\frac{2eg}{16\pi^2}\frac{3c_\mathrm{W}^2-s_\mathrm{W}^2}{c_\mathrm{W}}J_2(p^2,m_{\chi^{\pm\pm}}^2),\nonumber\\*
\\
\Pi_{ZZ}(p^2)&=&\frac{1}{16\pi^2}\bigg\{\sum_{i,j=1}^{3}\big[|g_{Z\chi_i^0\chi_j^0}|^2J_1(p^2,m_{\chi_i^0}^2,m_{\chi_j^0}^2)\nonumber\\
&&\qquad\qquad\qquad+m_{\chi_i^0}m_{\chi_j^0}(g_{Z\chi_i^0\chi_j^0}^2+\mathrm{h.c.})B_0(p^2,m_{\chi_i^0}^2,m_{\chi_j^0}^2)\big]\nonumber\\
&&\quad+\sum_{i,j=1}^3\big[(|a_{Z\chi_i^+\chi_j^-}|^2+|b_{Z\chi_i^+\chi_j^-}|^2)J_1(p^2,m_{\chi_i^\pm}^2,m_{\chi_j^\pm}^2)\nonumber\\
&&\qquad\qquad\quad-2m_{\chi_i^\pm}m_{\chi_j^\pm}(a_{Z\chi_j^+\chi_i^-}b_{Z\chi_i^+\chi_j^-}+\mathrm{h.c.})B_0(p^2,m_{\chi_i^\pm}^2,m_{\chi_j^\pm}^2)\big]\nonumber\\
&&\quad+\frac{g^2(3c_\mathrm{W}^2-s_\mathrm{W}^2)^2}{c_\mathrm{W}^2}J_2(p^2,m_{\chi^{\pm\pm}}^2)\bigg\},\\
\Pi_{WW}(p^2)&=&\frac{1}{16\pi^2}\bigg\{\sum_{i,j=1}^3\big[(|a_{W\chi_j^+\chi_i^0}|^2+|b_{W\chi_j^+\chi_i^0}|^2)J_1(p^2,m_{\chi_i^0}^2,m_{\chi_j^\pm}^2\nonumber)\\
&&\qquad\qquad\qquad-2m_{\chi_i^0}m_{\chi_j^\pm}(a_{W\chi_j^+\chi_i^0}b_{W\chi_j^+\chi_i^0}^\ast+\mathrm{h.c.})B_0(p^2,m_{\chi_i^0}^2,m_{\chi_j^\pm}^2)\big]\nonumber\\
&&\quad+\sum_{i=1}^3\big[(|a_{W\chi^{++}\chi_i^+}|^2+|b_{W\chi^{++}\chi_i^+}|^2)J_1(p^2,m_{\chi_i^\pm}^2,m_{\chi^{\pm\pm}}^2\nonumber)\\
&&\qquad\qquad-2m_{\chi^{\pm\pm}}m_{\chi_i^\pm}(a_{W\chi^{++}\chi_i^+}b_{W\chi^{++}\chi_i^+}^\ast+\mathrm{h.c.})B_0(p^2,m_{\chi_i^\pm}^2,m_{\chi^{\pm\pm}}^2)\big]\bigg\},\qquad\quad
\end{eqnarray}
where the related couplings are
\begin{eqnarray}
a_{Z\chi_i^+\chi_j^-}&=&gc_\mathrm{W}(\mathcal{C}_\mathrm{L})_{1,i}^\ast(\mathcal{C}_\mathrm{L})_{1,j}+\frac{g(3c_\mathrm{W}^2+s_\mathrm{W}^2)}{2c_\mathrm{W}}(\mathcal{C}_\mathrm{L})_{2,i}^\ast(\mathcal{C}_\mathrm{L})_{2,j}
+\frac{g(c_\mathrm{W}^2-s_\mathrm{W}^2)}{2c_\mathrm{W}}(\mathcal{C}_\mathrm{L})_{3,i}^\ast(\mathcal{C}_\mathrm{L})_{3,j},\nonumber\\*
\\
b_{Z\chi^+_i\chi^-_j}&=&gc_\mathrm{W}(\mathcal{C}_\mathrm{R})_{1,i}(\mathcal{C}_\mathrm{R})_{1,j}^\ast+\frac{g(c_\mathrm{W}^2-s_\mathrm{W}^2)}{2c_\mathrm{W}}(\mathcal{C}_\mathrm{R})_{2,i}(\mathcal{C}_\mathrm{R})_{2,j}^\ast
+\frac{g(3c_\mathrm{W}^2+s_\mathrm{W}^2)}{2c_\mathrm{W}}(\mathcal{C}_\mathrm{R})_{3,i}(\mathcal{C}_\mathrm{R})_{3,j}^\ast,\nonumber\\*
\\
g_{Z\chi^0_i\chi^0_j}&=&-\frac{g}{2c_\mathrm{W}}(\mathcal{N}_{2,i}\mathcal{N}_{2,j}^\ast-\mathcal{N}_{3,i}\mathcal{N}_{3,j}^\ast),\\
a_{W\chi_i^+\chi_j^0}&=&g(\mathcal{C}_\mathrm{L})_{1,i}^\ast\mathcal{N}_{1,j}+\frac{\sqrt{6}}{2}g(\mathcal{C}_\mathrm{L})_{2,i}^\ast\mathcal{N}_{2,j}+\sqrt{2}g(\mathcal{C}_\mathrm{L})_{3,i}^\ast\mathcal{N}_{3,j},\\
b_{W\chi_i^+\chi_j^0}&=&g(\mathcal{C}_\mathrm{R})_{1,i}\mathcal{N}_{1,j}^\ast-\sqrt{2}g(\mathcal{C}_\mathrm{R})_{2,i}\mathcal{N}_{2,j}^\ast-\frac{\sqrt{6}}{2}g(\mathcal{C}_\mathrm{R})_{3,i}\mathcal{N}_{3,j}^\ast.
\end{eqnarray}

Similar to the SDFDM and DTFDM models, $y_1=\pm y_2$ leads to the custodial symmetry, and hence $T=U=g_{Z\chi_1^0\chi_1^0}=0$. For $y_1=y_2=y$, $\mathrm{SU}(2)_\mathrm{R}$ doublets
\begin{equation}
(\mathcal{Q}^A)^{ij}_k=\begin{pmatrix}(Q_1)_k^{ij}\\ (Q_2)_k^{ij}\end{pmatrix},\quad
(\mathcal{H}^A)^i=\begin{pmatrix}H_i^\dag\\ H_i\end{pmatrix},
\end{equation}
can be used to manifest $\mathrm{SU}(2)_\mathrm{L}\times\mathrm{SU}(2)_\mathrm{R}$ invariant Lagrangians
\begin{eqnarray}
\mathcal{L}_\mathrm{Q}&=&i\mathcal{Q}_A^\dag\bar{\sigma}^\mu D_\mu\mathcal{Q}^A-\frac{1}{2}[m_Q\epsilon_{AB}\epsilon_{il}(\mathcal{Q}^A)_k^{ij}(\mathcal{Q}^B)_j^{lk}+\mathrm{h.c.}],\\
\mathcal{L}_\mathrm{HTQ}&=&y\epsilon_{AB}(\mathcal{Q}^A)^{jk}_iT^i_k(\mathcal{H}^B)_j+\mathrm{h.c.}.
\end{eqnarray}
In this case, $g_{h\chi_1^0\chi_1^0}=0$ holds for $m_Q<m_T$, leading to a vanishing SI DM-nucleon scattering cross section at tree level.

\subsection{Expected Constraints}

\begin{figure}[!t]
\centering
\subfigure[~$m_Q<m_T$ case\label{TQSTU:a}]
{\includegraphics[width=0.49\textwidth]{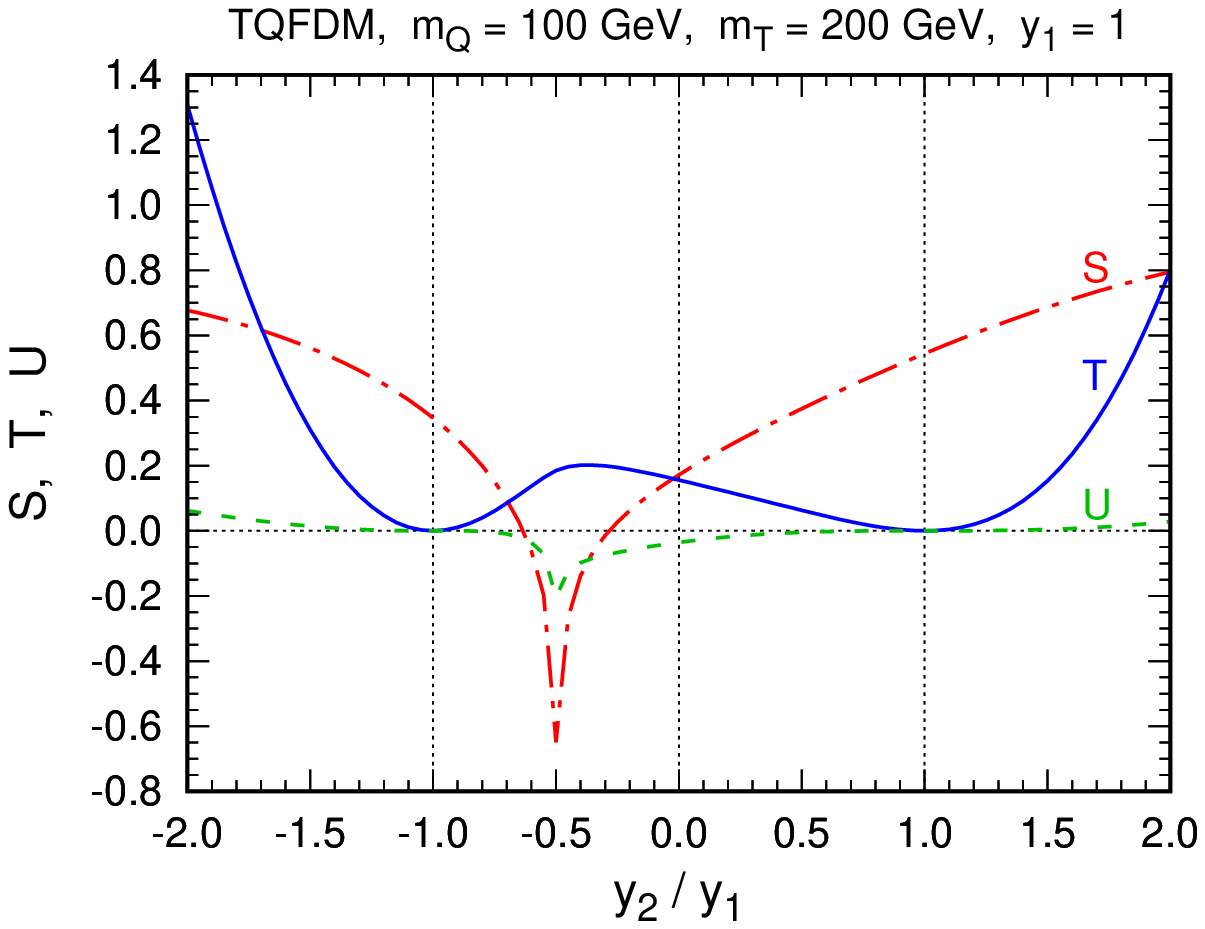}}
\subfigure[~$m_Q>m_T$ case\label{TQSTU:b}]
{\includegraphics[width=0.49\textwidth]{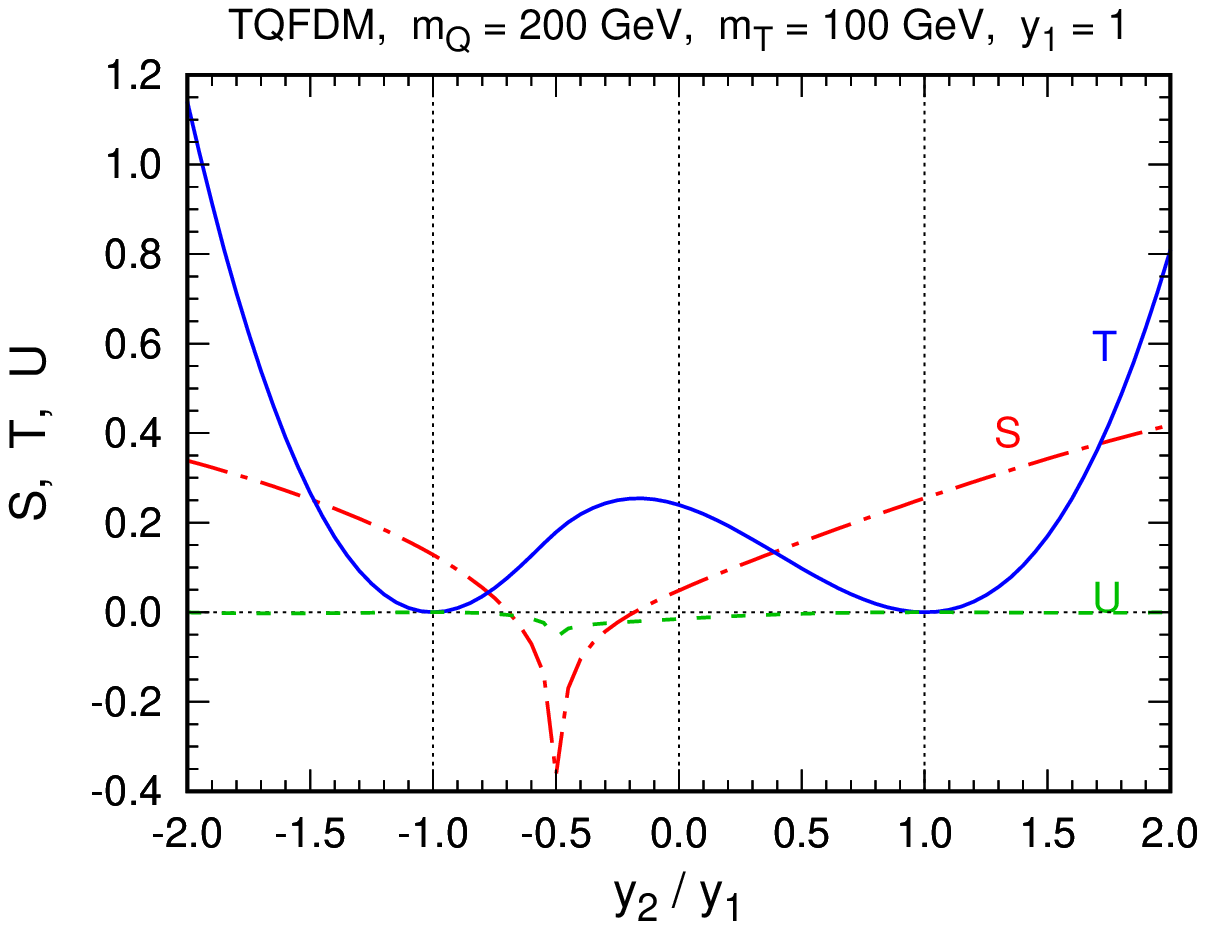}}
\caption{$S$, $T$, and $U$ as functions of $y_2/y_1$ in the TQFDM model  with $y_1=1$. In the left (right) panel, $m_Q=100~(200)~\si{GeV}$ and  $m_T=200~(100)~\si{GeV}$.}
\label{TQSTU}
\end{figure}

We demonstrate the behaviors of EW oblique parameters as functions of $y_2/y_1$ for the TQFDM model with fixed mass parameters in Fig.~\ref{TQSTU}.
For $y_2/y_1=\pm 1$, $T$ and $U$ arrive at zero, due to the custodial symmetry.
The $S$ parameter has a dip at $y_2/y_1\approx-0.5$, where $m_{\chi_1^0}$ and $m_{\chi_1^\pm}$ approach to zero, leading to large contributions to vacuum polarizations.

\begin{figure}[!t]
\centering
\subfigure[~$y_1=y_2=1$ (custodial symmetry)\label{TQS2d:a}]
{\includegraphics[width=0.49\textwidth]{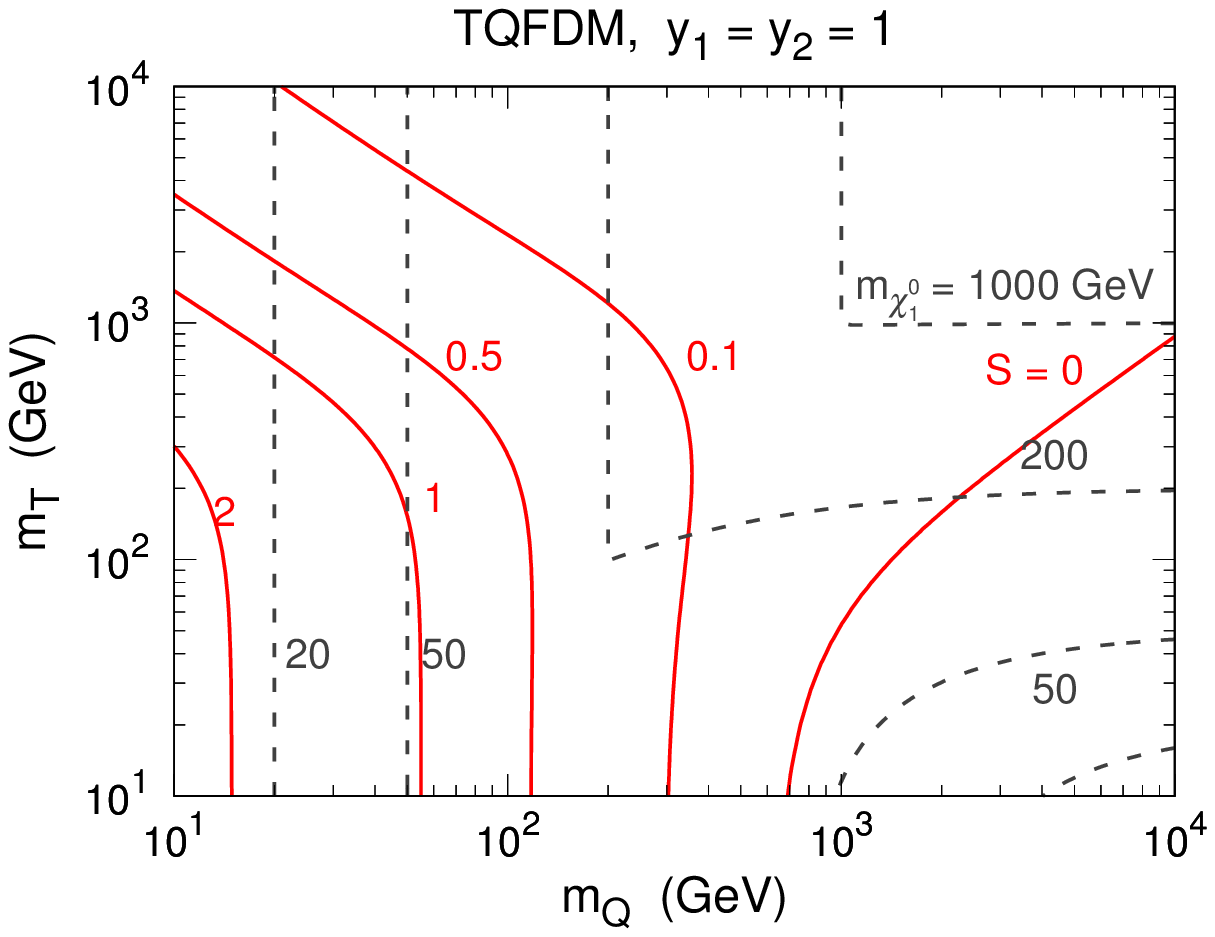}}
\subfigure[~$y_1=1$, $y_2=1.5$\label{TQS2d:b}]
{\includegraphics[width=0.49\textwidth]{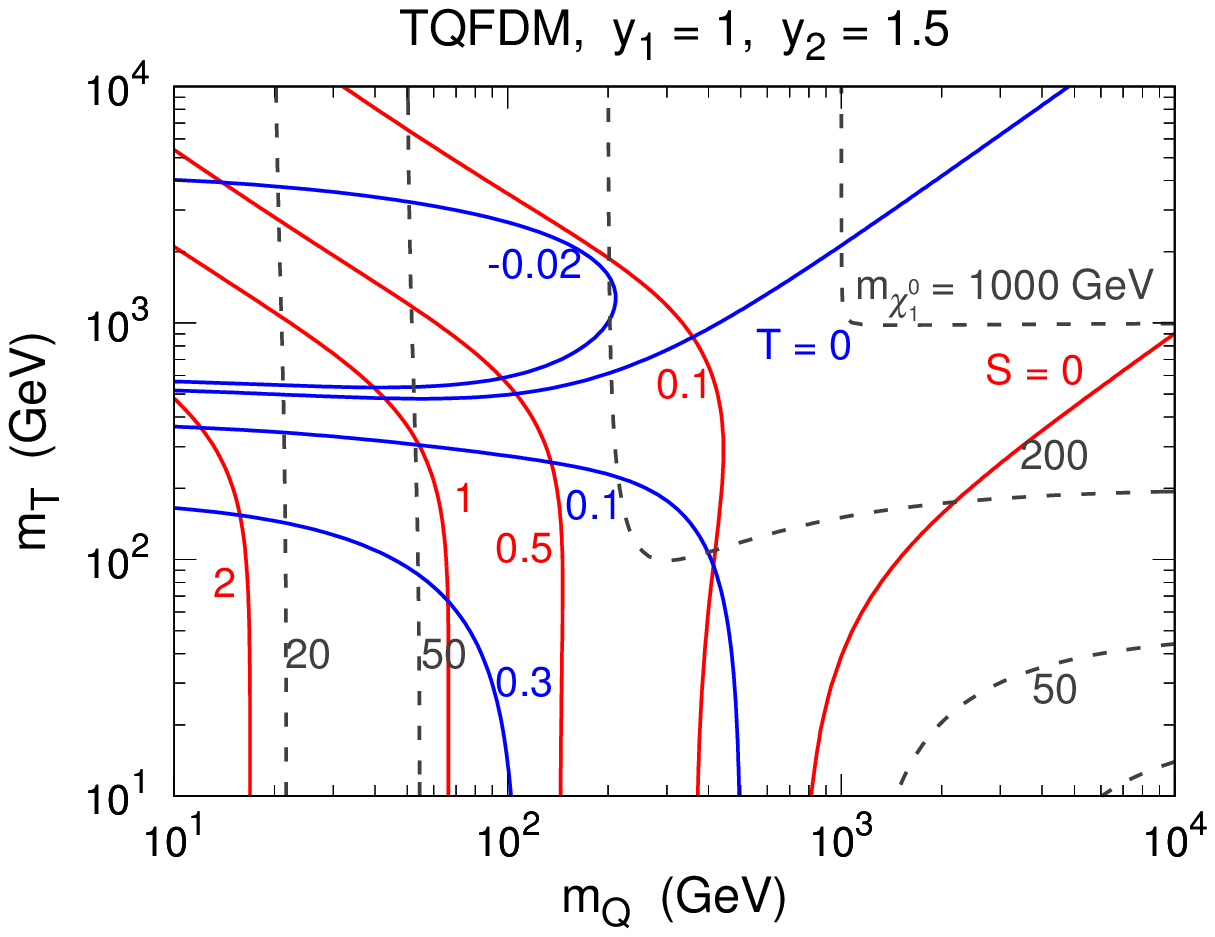}}
\caption{Contours of $S$ (red solid lines), $T$ (blue solid lines), and $m_{\chi_1^0}$ (gray dashed lines) in the $m_Q-m_T$ plane for the TQFDM model with fixed $y_1$ and $y_2$.}
\label{TQS2d}
\end{figure}

Fig.~\ref{TQS2d} exhibits the contours of $S$ and $T$ in the $m_Q-m_T$ plane with fixed Yukawa couplings.
The behaviors of $S$ and $T$ are quite similar to those in the SDFDM model, but the values of $S$ are larger, since the gauge interactions are stronger.
For the custodial symmetry limit $y_1=y_2=1$, which corresponds to Fig.~\ref{TQS2d:a}, we can have an approximate analysis on $S$, analogous to that in Subsection~\ref{subsec:SDFDM:cstr}.
When $m_Q<m_T\ll yv$, the mass spectrum is $m_{\chi_1^0}= m_{\chi_1^\pm}=m_{\chi^{\pm\pm}} = m_Q$, $m_{\chi_{2,3}^0}\approx m_{\chi_{2,3}^\pm}\approx \sqrt{6}yv/3$, resulting in a significant $\ln(yv/m_Q)$ term for $S$.
We can conclude that the similar behaviors of $S$ in the SDFDM and TQFDM models is because there is an unmixed particle, either $\chi^\pm$ or $\chi^{\pm\pm}$.
In contrast, dark sector fermions are all mixed with each other in the DTFDM model, producing a very different behavior.
For $y_1=1$ and $y_2=1.5$, which corresponds to Fig.~\ref{TQS2d:b}, not only the behavior of $T$ but also the values are analogous to those in the SDFDM model.

\begin{figure}[!t]
\centering
\subfigure[~$y_1=y_2=1$ (custodial symmetry)\label{TQ2d:a}]
{\includegraphics[width=0.49\textwidth]{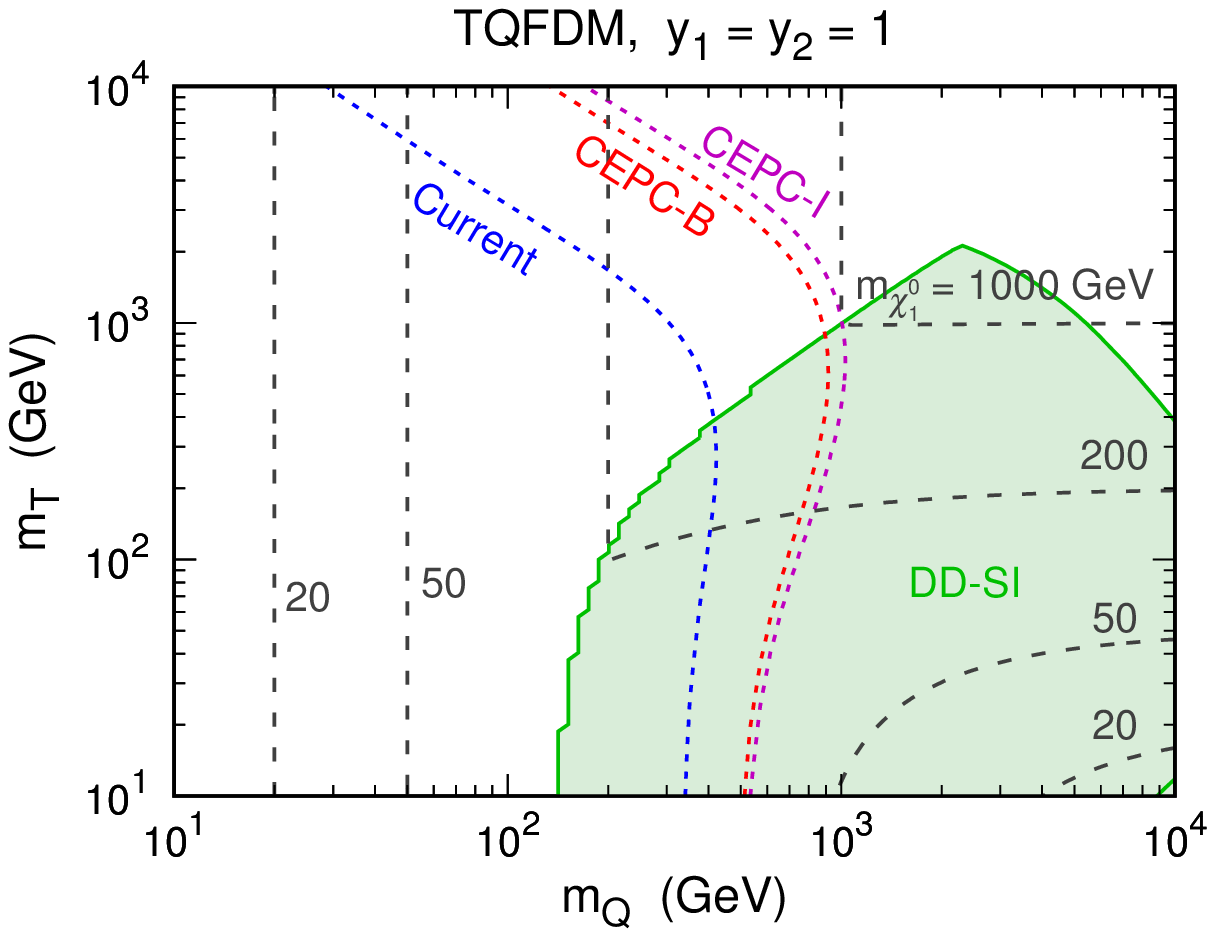}}
\subfigure[~$y_1=1$, $y_2=1.5$\label{TQ2d:b}]
{\includegraphics[width=0.49\textwidth]{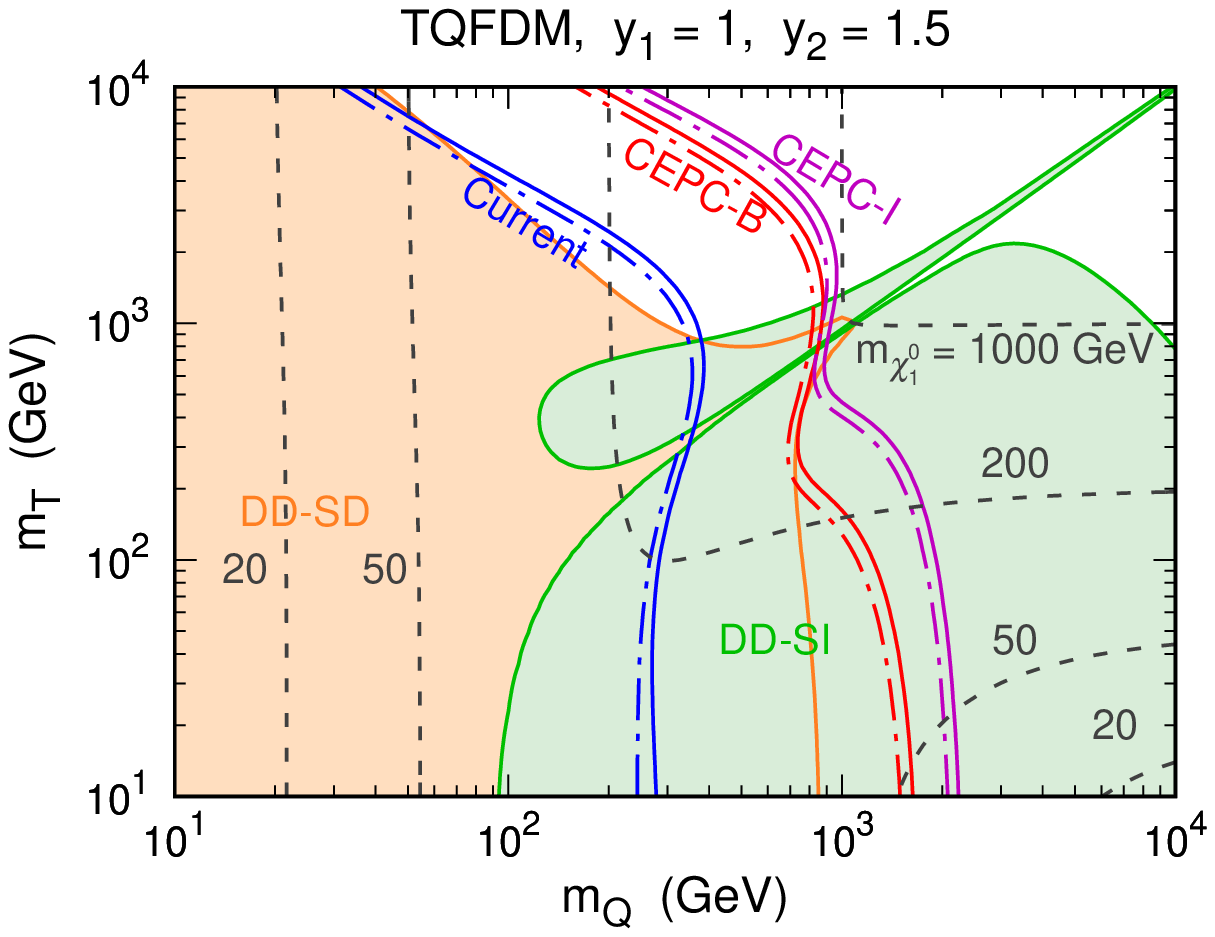}}
\caption{Expected 95\% CL constraints from CEPC precisions of EW oblique parameters in the $m_Q-m_T$ plane for the TQFDM model with fixed $y_1$ and $y_2$. The meanings of labels, colors, and line types are the same as in Fig.~\ref{SD2d}.}
\label{TQ2d}
\end{figure}

In Fig.~\ref{TQ2d}, we show both the expected constraints from oblique parameters and the current direct detection bounds in the $m_Q-m_T$ plane.
The exclusion regions from direct detection are quite analogous to those in the SDFDM and DTFDM models, since the intrinsic physics is basically identical.
Moreover, the limits from EW precision measurements have similar behaviors to those in the SDFDM model, due to the reason discussed above.
But the expected exclusion regions are enlarged.
For both the case of $y_1=y_2=1$ in Fig.~\ref{TQ2d:a} and the case of $y_1=1$ and $y_2=1.5$ in Fig.~\ref{TQ2d:b}, CEPC EW data could explore up to $m_{\chi_1^0}\sim 1~\si{TeV}$.

\section{Conclusions and Discussions}
\label{sec:concl}

The future CEPC project will greatly improve EW precision measurements, leading to an unprecedented precision of EW oblique parameters.
This will provide an excellent opportunity to indirectly test new physics with EW interactions, in particular, WIMP dark matter.
In this work, we calculate the expected constraints from CEPC EW data on fermionic WIMP dark matter.
Current direct detection bounds are also demonstrated for comparison.

The expected CEPC precisions of oblique parameters are derived through global fits assuming reduction of the uncertainties of EW precision observables due to future CEPC data and theoretical efforts.
Fit results are obtained for the case where all the oblique parameters are free and for the cases assuming some of them vanish.
We have used these results to study the CEPC sensitivity to three WIMP models, \textit{i.e.}, the SDFDM, DTFDM, and TQFDM models.

Each of these models has a dark sector consisting of fermionic multiplets in two $\mathrm{SU}(2)_\mathrm{L}$ representations whose dimensions differ by one and allow two kinds of Yukawa couplings to the SM Higgs doublet. The DM candidate $\chi_1^0$ is the lightest mass eigenstate of multiplet neutral components.
When the two Yukawa couplings are equal, there is a custodial symmetry resulting in vanishing DM couplings to the Higgs and $Z$ bosons in a particular region of the parameter space.
In this case, direct detection experiments can hardly probe the model, while CEPC EW data would still be very sensitive.
Moreover, in the case with custodial symmetry violation, CEPC can also explore further than current direct detection.
In some moderate values of Yukawa couplings, we find that CEPC data are expected to probe up to $m_{\chi_1^0}\sim 600~\si{GeV}$, $2~\si{TeV}$, and $1~\si{TeV}$ in the SDFDM, DTFDM, and TQFDM models, respectively.

LHC searches for production of dark sector fermions are also important for studying these models.
Nevertheless, the LHC sensitivity is limited by the low electroweak production rates and complicated final states.
Since CEPC EW data can reach up to TeV mass scales, as shown above, the CEPC sensitivity could be much better than LHC.
It is worth emphasizing that collider studies on dark matter are free from astrophysical and cosmological factors, only depending on its properties in particle physics.

In contrast, the interpretations of direct and indirect detection experimental results depend on many astrophysical inputs, \textit{e.g.}, the local DM density, $J$-factors of dwarf galaxies, and ambiguous astrophysical backgrounds.
The information inferred from the observed DM relic abundance is not totally solid, since the calculation may be affected by nonstandard cosmological evolution.
Therefore, collider studies should be treated as an independent and robust way for exploring DM particle nature.

\acknowledgments

We thank Bin Zhu, Dan-Yang Liu, and Zhong-Hui Zhang for discussions.
This work is supported by the National Natural Science Foundation of China (NSFC)
under Grant Nos. 11375277, 11410301005, 11647606 and 11005163, the Fundamental Research
Funds for the Central Universities,
the Natural Science Foundation of Guangdong Province under Grant No. 2016A030313313,
and the Sun Yat-Sen University Science Foundation.
ZHY is supported by the Australian Research Council.

\appendix

\section{Dark Matter Scattering off Nucleons}
\label{app:DM_DD}

In the SDFDM, DTFDM, and TQFDM models, the DM candidate $\chi_1^0$ may have nonzero couplings to the Higgs and $Z$ bosons. The exchange of a Higgs boson between $\chi_1^0$ and nuclei leads to SI scattering, while the exchange of a $Z$ boson leads to SD scattering. Therefore, direct detection experiments have potential to explore these models. In this appendix, we provide the expressions for calculating the scattering cross sections.

The Lagrangian for the trilinear interaction between the Majorana fermion $\chi_1^0$ and the Higgs boson is given by Eq.~\eqref{eq:L_hchichi}.
For zero momentum transfer, it induces an effective operator describing the scalar interaction between $\chi_1^0$ and a nucleon $N$:
\begin{equation}
\mathcal{L}_{\mathrm{S},N}=\sum_{N=p,n}G_{\mathrm{S},N}\bar{\chi}_1^0\chi_1^0\bar{N}N,
\end{equation}
with
\begin{equation}
G_{\mathrm{S},N}=-\frac{g_{h\chi_1^0\chi_1^0}m_N}{2vm_h^2}\left(\sum_{q=u,d,s}f_q^N+3f_Q^N\right).
\end{equation}
The nucleon form factors $f_i^N$ are given by~\cite{Ellis:2000ds}
\begin{eqnarray}
&&f_u^p=0.020\pm0.004,\quad f_d^p=0.026\pm0.005,\quad f_u^n=0.014\pm0.003,\\
&&f_d^n=0.036\pm0.008,\quad f_s^p=f_s^n=0.118\pm0.062,\quad f_Q^N=\frac{2}{27}\left(1-\sum_{q=u,d,s}f_q^N\right).\qquad
\end{eqnarray}
The SI scattering cross section due to this effective interaction can be expressed as~\cite{Zheng:2010js}
\begin{equation}
\sigma_{\chi N}^\mathrm{SI}=\frac{4}{\pi}\mu_{\chi N}^2 G_{\mathrm{S},N}^2,
\end{equation}
where $\mu_{\chi N}\equiv m_{\chi_1^0}m_N/(m_{\chi_1^0}+m_N)$ is the reduced mass.

The Lagrangian for the trilinear interaction between $\chi_1^0$ and the $Z$ boson is given by Eq.~\eqref{eq:L_Zchichi}.
It leads to an effective operator for the axial vector interaction:
\begin{equation}
\mathcal{L}_{\mathrm{A},N}=\sum_{N=p,n}G_{\mathrm{A},N}\bar{\chi}_1^0\gamma^\mu\gamma_5\chi_1^0\bar{N}\gamma_\mu\gamma_5N,
\end{equation}
with
\begin{equation}
G_{\mathrm{A},N}=\sum_{q=u,d,s}G_{\mathrm{A},q}\Delta_q^N,\quad
G_{\mathrm{A},q}=\frac{gg^q_\mathrm{A}g_{Z\chi_1^0\chi_1^0}}{4c_\mathrm{W}m_Z^2},
\end{equation}
where $g_\mathrm{A}^u=1/2$, $g_\mathrm{A}^d=g_\mathrm{A}^s=-1/2$, and the form factors are~\cite{Airapetian:2006vy}
\begin{equation}
\Delta_u^p=\Delta_d^n=0.842\pm0.012,\quad \Delta_d^p=\Delta_u^n=-0.427\pm0.013,\quad \Delta_s^p=\Delta_s^n=-0.085\pm0.018.
\end{equation}
This effective interaction induces SD scattering with a cross section~\cite{Zheng:2010js}
\begin{equation}
\sigma_{\chi N}^\mathrm{SD}=\frac{12}{\pi}\mu_{\chi N}^2 G_{\mathrm{A},N}^2.
\end{equation}

\bibliographystyle{JHEP}
\bibliography{FDM}

\end{document}